\documentclass[aps,pra,twocolumn,amsmath,amssymb,superscriptaddress,10pt,nobibnotes]{revtex4-2}

\usepackage{amsfonts, amssymb, amsmath, dsfont,mathrsfs}
\usepackage{graphicx}
\usepackage{color}
\usepackage[utf8]{inputenc}
\usepackage[T1]{fontenc}
\usepackage[svgnames]{xcolor}
\usepackage{textcomp}

\usepackage{array}
\newcommand{\PreserveBackslash}[1]{\let\temp=\\#1\let\\=\temp}
\newcolumntype{C}[1]{>{\PreserveBackslash\centering}p{#1}}
\newcolumntype{R}[1]{>{\PreserveBackslash\raggedleft}p{#1}}
\newcolumntype{L}[1]{>{\PreserveBackslash\raggedright}p{#1}}

\usepackage{hyperref}
\hypersetup{colorlinks,urlcolor=DarkBlue,linkcolor=DarkBlue,citecolor=DarkBlue,pdfdisplaydoctitle=true,pdfpagemode=UseOutlines,bookmarksnumbered=true,bookmarksopen=true} 

\newcommand{\bra}{\langle}
\newcommand{\ket}{\rangle}

\newcommand{\out}{\mathrm{out}}
\newcommand{\ins}{\mathrm{in}}
\newcommand{\rf}{\mathrm{ref}}

\newcommand{\Ve}{\operatorname{Var}}

\newcommand{\re}{\operatorname{Re}}
\newcommand{\im}{\operatorname{Im}}
\newcommand{\tr}{\operatorname{Tr}}

\newcommand{\erfc}{\operatorname{erfc}}

\newcommand{\eq}[1]{Eq.~\eqref{#1}}

\newcommand{\Eq}[1]{Equation~\eqref{#1}}

\newcommand{\fig}[1]{Fig.~\ref{#1}}

\newcommand{\tab}[1]{Table~\ref{#1}}

\newcommand{\beginsupplement}{%
	\setlength{\parskip}{3pt}
	\setcounter{table}{0}
	\renewcommand{\thetable}{S\arabic{table}}%
	\setcounter{figure}{0}
	\renewcommand{\thefigure}{S\arabic{figure}}%
	\setcounter{section}{0}
	\renewcommand{\thesection}{S\arabic{section}}
	\setcounter{subsection}{0}
	\renewcommand{\thesubsection}{S\arabic{section}.\arabic{subsection}}
	\setcounter{equation}{0}
	\renewcommand{\theequation}{S\arabic{equation}}
}

\begin{document}
	
% =====================================================================================
% title
% =====================================================================================

\title{Optimal control of coherent light scattering for binary decision problems}
\author{Dorian Bouchet}
\altaffiliation{\href{mailto:dorian.bouchet@univ-grenoble-alpes.fr}{dorian.bouchet@univ-grenoble-alpes.fr}}
\affiliation{Université Grenoble Alpes, CNRS, LIPhy, 38000 Grenoble, France}
\author{Lukas M.\ Rachbauer}
\affiliation{Institute for Theoretical Physics, Vienna University of Technology (TU Wien), 1040 Vienna, Austria}
\author{Stefan Rotter}
\affiliation{Institute for Theoretical Physics, Vienna University of Technology (TU Wien), 1040 Vienna, Austria}
\author{Allard P.\ Mosk}
\affiliation{Nanophotonics, Debye Institute for Nanomaterials Science and Center for Extreme Matter and Emergent Phenomena, Utrecht University, P.O. Box 80000, 3508 TA Utrecht, Netherlands}
\author{Emmanuel Bossy}
\affiliation{Université Grenoble Alpes, CNRS, LIPhy, 38000 Grenoble, France}

% =====================================================================================
% abstract
% =====================================================================================

\begin{abstract}
Due to quantum noise fluctuations, the rate of error achievable in decision problems involving several possible configurations of a scattering system is subject to a fundamental limit known as the Helstrom bound. Here, we present a general framework to calculate and minimize this bound using coherent probe fields with tailored spatial distributions. As an example, we experimentally study a target located in between two disordered scattering media. We first show that the optimal field distribution can be directly identified using a general approach based on scattering matrix measurements. We then demonstrate that this optimal light field successfully probes the presence of the target with a number of photons that is reduced by more than two orders of magnitude as compared to unoptimized fields.
\end{abstract}

\maketitle

% =====================================================================================
% intro
% =====================================================================================
	
Many sensing applications rely on the detection of targets embedded within disordered or engineered materials. For instance, interferometric techniques are currently being developed to detect the presence of single particles within biological specimens based on coherent light scattering~\cite{taylor_interferometric_2019,young_interferometric_2019}. Coherent beams are also used to detect the presence of defects in nanofabricated samples such as integrated circuits~\cite{orji_metrology_2018,brown_machine_2020}. In such experiments, and more generally for all decision problems involving several possible configurations of a given scattering system, the rate of error is fundamentally limited by quantum noise fluctuations ~\cite{helstrom_quantum_1976,weedbrook_gaussian_2012}, which usually appear in measured data as shot noise. This limit, which predicts high rates of error for measurements performed in low-light conditions, represents a central obstacle for the development of non-destructive high-speed sensing techniques. Different strategies have been devised to address this challenge, notably by finding optimal and robust receivers for coherent states \cite{cook_optical_2007,weedbrook_gaussian_2012,becerra_experimental_2013,sych_practical_2016,solis-prosser_experimental_2017,dimario_robust_2018} and by adopting quantum illumination schemes~\cite{lloyd_enhanced_2008,pirandola_advances_2018,nair_fundamental_2020}; however, all these approaches are primarily applied to scattering systems with simple permittivity distributions.

When interacting with complex media such as disordered media or engineered nano-materials, light is usually absorbed and scattered multiple times, resulting in the formation of complex interference patterns. Despite this difficulty, it has been shown that the propagation of light in such media can be controlled by spatially modulating the incident field using wavefront shaping protocols~\cite{mosk_controlling_2012,horstmeyer_guidestar-assisted_2015,rotter_light_2017}. These methods have opened up the possibility to optimally deposit and store energy in or behind scattering materials~\cite{vellekoop_universal_2008,kim_maximal_2012,popoff_coherent_2014,cheng_focusing_2014,ambichl_focusing_2017,durand_optimizing_2019,bender_depth-targeted_2021}. Integrated in a framework based on estimation theory, wavefront shaping also enables the generation of optimal fields that maximize the Fisher information retrieved from complex scattering systems, thus allowing to precisely estimate small variations in the value of continuous parameters~\cite{bouchet_influence_2020,bouchet_maximum_2021}. The question then arises as to how wavefront shaping can be used for target detection and, more generally, for any decision problem in which a number of hypotheses are formulated and an experiment is conducted to decide on which hypothesis is true. This question is especially important in the development of optical microscopes optimized for (non-imaging) decision tasks, as promoted by the advent of approaches based on deep learning algorithms~\cite{hershko_multicolor_2019,kellman_physics-based_2019,muthumbi_learned_2019,hougne_learned_2020,horisaki_deeply_2020}.

In this Letter, we identify and experimentally generate coherent light fields that are optimally shaped for decision problems involving two possible configurations of a complex scattering system. To this end, we introduce the \textit{discrimination operator}, which allows one to directly identify the spatial distribution of the field that minimizes the rate of error due to quantum noise fluctuations. We first show that this operator can be readily constructed from the knowledge of the scattering matrices describing each configuration of the system. We then illustrate this approach experimentally by generating a light field that is optimized to detect the presence of a target hidden in between two disordered scattering media. This field is shown to optimally interact with the target, despite its complex environment. Finally, we demonstrate that the optimal field successfully probes the presence of the hidden target in low-light conditions, with a number of photons that is reduced by more than two orders of magnitude as compared to unoptimized fields. These results, which connect quantum detection theory to wavefront shaping, establish a new benchmark to assess and improve the performances of sensing and classification techniques using structured illumination. 

% =====================================================================================
% main text
% =====================================================================================

\begin{figure*}[t]
	\begin{center}
		\includegraphics[width=\linewidth]{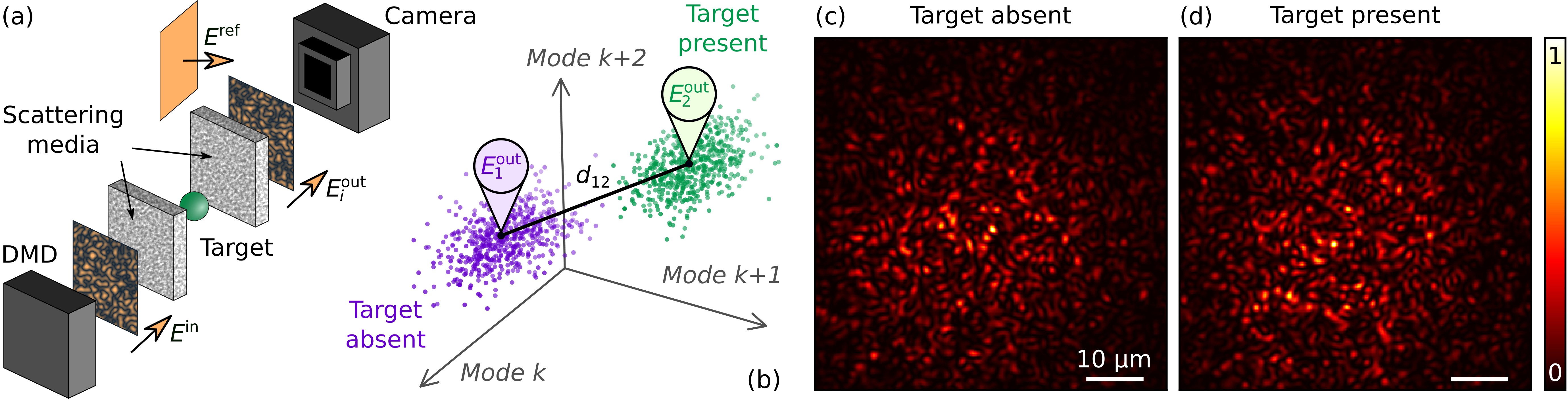}
	\end{center}
	\caption{(a) Representation of the experiment, which consists in optically probing the presence of a target ($3$\,\textmu m in diameter) located between two scattering media. The system is illuminated with an incident field that is spatially modulated using a digital micromirror device (DMD). The field that comes out of the system is measured by a camera in $N$ spatial modes using a homodyne scheme. (b)~The measured field spans a complex $N$-dimensional space. The expectation value of the field depends on the presence of the target, and its variance is ultimately limited by quantum noise fluctuations. The incident field optimally probes the presence of the target when the statistical distance $d_{12}$ is maximized. (c,d)~Intensity distributions experimentally measured with the optimal incident field when the hidden target is (c)~absent and (d)~present. Due to complex absorption and scattering processes involved within the system, these distributions appear as speckle patterns. Despite this complexity, the optimal light field is strongly affected by the presence of the target, which allows us to detect it with a minimum rate of error. }
	\label{fig1}
\end{figure*}

We consider an arbitrarily complex scattering system that can take two distinct configurations with probabilities $\pi_1$ and $\pi_2$. Detecting the presence of a target included within a given scattering medium (\fig{fig1}a) constitutes a typical example of such a situation: the target can either be absent (hypothesis $H_1$) or present (hypothesis $H_2$). To decide on which hypothesis is true, we apply a measurement to the scattering system and we choose a hypothesis based on a decision criterion, resulting in a probability of error $P_{\mathrm{err}}$. This probability can be minimized by optimizing over both the decision criterion and the positive operator-valued measure (POVM) describing the measurement process. The minimum probability of error $P_{\mathrm{H}}$, which is limited only by quantum noise fluctuations, is calculated from the trace distance between the two quantum states associated with each hypothesis---a result known as the Helstrom bound~\cite{helstrom_quantum_1976,weedbrook_gaussian_2012}. This general formalism can be applied to the case of a probe field in a coherent state described by the coefficients $\lbrace E^\ins_1,\dots,E^\ins_M \rbrace$ in $M$ spatial modes. After interacting with the scattering system, such an incident state produces an outgoing state that is similarly described by the coefficients $\lbrace E^\out_{i,1},\dots,E^\out_{i,N} \rbrace$ in $N$ spatial modes, where the sub-index $i$ denotes the configuration of the scattering system interacting with the field (if $H_1$ is true $i=1$, and if $H_2$ is true $i=2$). Note that these coefficients, defined here as expectation values (calculated over quantum noise fluctuations) of the complex field operator, can equivalently be interpreted as describing the complex classical field. The Helstrom bound is then simply expressed by~\cite{helstrom_quantum_1976}
\begin{equation}
P_\mathrm{H} = \frac{1}{2} \left( 1- \sqrt{1- 4 \pi_1 \pi_2 \exp \left( - n d_{12}^2 \right)} \right) ,
\label{eq:error_helstrom}
\end{equation}
where $n$ is the number of incident photons and $d_{12}$ is a statistical distance expressed by
\begin{equation}
d_{12}^2 = \frac{1}{n} \sum_{k=1}^N \left \vert E_{2,k}^\out - E_{1,k}^\out \right \vert ^2 .
\label{eq:distance}
\end{equation}
The distance $d_{12}$ quantifies the overlap between two different outgoing states (\fig{fig1}b), as created by the interaction of a single photon with each of the two possible configurations of the scattering system. This distance consequently drives the exponential decay of the probability of error that is achieved when both the POVM and the decision criterion are optimized. Indeed, in the asymptotic limit ($n \gg d_{12}^{-2}$), the Helstrom bound decays exponentially with $n$, with a decay constant given by $d_{12}^{2}$. \Eq{eq:distance} also explicitly shows that, for a binary decision problem, the contribution of each outgoing mode simply sums up to set the Helstrom bound.

\begin{figure*}[t]
	\begin{center}
		\includegraphics[width=\linewidth]{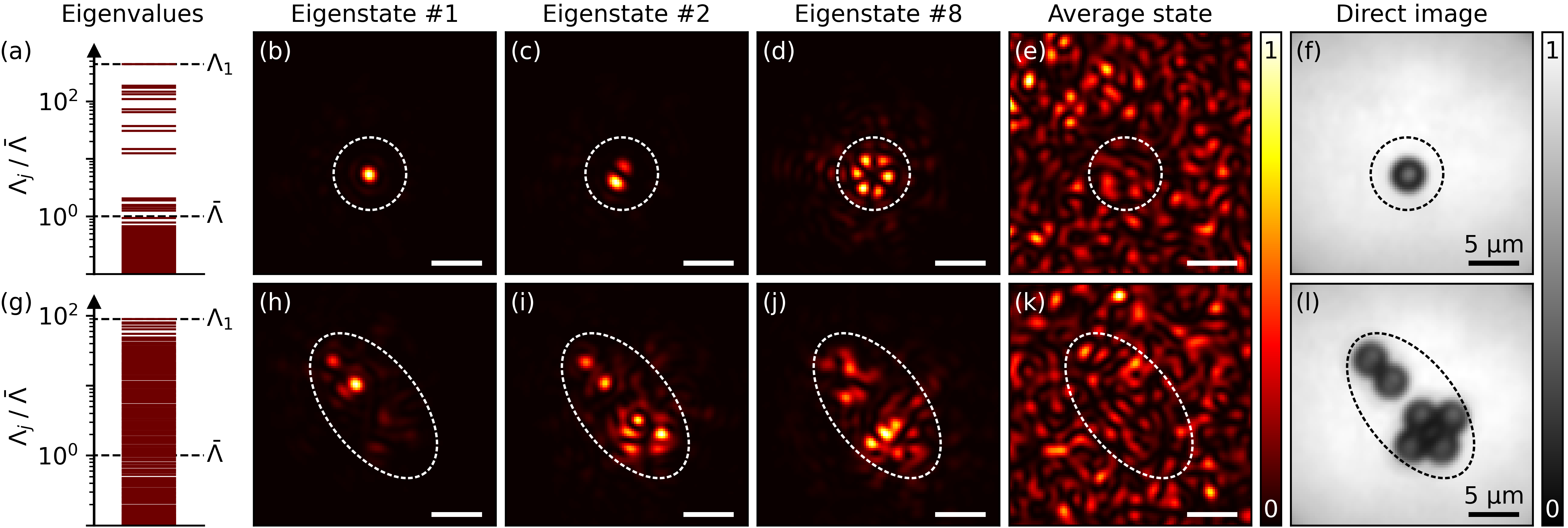}
	\end{center}
	\caption{(a)~Eigenvalues $\Lambda_j$ of the discrimination operator, normalized by the average value $\bar{\Lambda}$ (the averaging is performed over all possible incident states). Small eigenvalues (associated with measurement noise) are closely spaced and appear as a continuum due to the finite thickness of the lines, while large eigenvalues (associated with eigenstates that significantly interact with the target) appear as discrete lines. (b-e)~Intensity distribution in the target plane measured for a few representative eigenstates---including (b)~the first eigenstate (i.e. the optimal state), (c)~the second eigenstate, and (d)~the eighth eigenstate---as well as for (e)~the average state (defined as an equally-weighted linear superposition of all eigenstates). These distributions were measured in the absence of the target, and without the scattering medium located between the target and the camera. (f)~Image of the target (a single bead) measured under spatially-incoherent illumination. (g-l)~Analogous to (a-f) for a target composed of six beads.}
	\label{fig2}
\end{figure*}

Far-field wavefront shaping techniques enable us to generate incident states with custom spatial distributions~\cite{mosk_controlling_2012}. Here, our goal is to identify the optimal field distribution that minimizes the Helstrom bound for any given number $n$ of incident photons, which entails maximizing $d_{12}^2$ over all possible incident states. For this purpose, we introduce the scattering matrices $S_1$ and $S_2$~\cite{rotter_light_2017}, which connect incident to outgoing states under the respective hypotheses (target being absent or present). These scattering matrices are supposed to be known, either by \textit{ab initio} calculations or by prior measurements. Conveniently writing \eq{eq:distance} in bra-ket notation $ n d_{12}^2= \bra E_{2}^\out - E_{1}^\out | E_{2}^\out - E_{1}^\out \ket $, and introducing the linear relation defining the $S$-matrix $| E^\out_i \ket= S_i | E^\ins \ket$, we obtain the following quadratic form (see~\cite{supplementary}, Section~S1.1):
\begin{equation}
	d_{12}^2 = \bra \mathcal{E}^\ins | D_{12} | \mathcal{E}^\ins \ket ,
	\label{eq:dist}
\end{equation}
where $ | \mathcal{E}^\ins \ket = n^{-1/2} |E^\ins \ket$ is the normalized incident state and where $D_{12} = (S_2-S_1)^\dagger (S_2-S_1) $ is a Hermitian operator that we refer to as the \textit{discrimination operator}. Among the $M$ eigenvalues of $D_{12}$, which all lie in the interval $[0;4]$, the largest one is of specific interest as it gives the maximum achievable value of $d_{12}^2$, which is reached by illuminating the scattering system with the corresponding eigenstate. This general approach thus allows one to minimize the Helstrom bound from the knowledge of the scattering matrices $S_1$ and $S_2$, regardless of the complexity of absorption and scattering processes that are involved within the system (\fig{fig1}c,d).

In order to investigate the properties of the discrimination operator experimentally, we study a target (a polystyrene bead with a diameter of $3$\,\textmu m) located on a glass coverslip, that we place in between ground glass diffusers (\fig{fig1}a, see also~\cite{supplementary} Sections~S3 and~S4). This scattering system is illuminated using coherent light at a wavelength $\lambda=532$\,nm. To characterize the system through transmission matrix measurements~\cite{popoff_measuring_2010}, we modulate the amplitude and the phase of the incident field with a digital micromirror device (DMD) using Lee holography~\cite{lee_binary_1974}, and we measure both quadratures of the outgoing field with a camera using off-axis holography~\cite{cuche_spatial_2000}. In this way, we control $M=1735$~incident modes and $N=2617$~outgoing modes, allowing us to acquire two (sub-unitary) transmission matrices $S_1$ and $S_2$, measured respectively with the target present and absent from the field of view. 

The knowledge of these matrices allows us to construct the discrimination operator $D_{12}$, and thus to identify the optimal incident state via an eigenvalue decomposition of $D_{12}$. To quantitatively assess the benefits of operating with the optimal state instead of unoptimized ones, we can study the $M$ eigenvalues of $D_{12}$, which we sort in descending order (\fig{fig2}a). Using this convention, the maximal value of $d_{12}^2$ is given by the first eigenvalue $\Lambda_1$, which we compare to the value of $d_{12}^2$ averaged over all possible incident states. This average value $\bar{\Lambda}=\tr (D_{12})/M$ is exactly reached for an equally-weighted linear superposition of all eigenstates, that we refer to as the average state. In our experiment, using the optimal state instead of the average one is shown to drastically enhance $d_{12}^2$, with a ratio $\Lambda_1/\bar{\Lambda}$ of the order of $400$. This implies that the number of incident photons needed to reach a given probability of error is smaller by more than two orders of magnitude with the optimal state. 

In order to acquire a better understanding of the field distribution within the system, we kept the first diffuser (located between the DMD and the target) but temporarily removed the second diffuser (located between the target and the camera). In this configuration, the camera directly images the plane of the target, thereby allowing us to visualize how the different eigenstates interact with the target (\fig{fig2}b-f). We can see that the first eigenstate optimally interacts with the target by generating a strong focus at its position. Remarkably, the intensity distribution of all eigenstates that strongly interact with the target (those associated with the largest eigenvalues) have a structured aspect that resemble those of Laguerre-Gaussian modes (\fig{fig2}b-d, see also~\cite{supplementary} Section~S5.1), as opposed to the speckle-like distribution generated by the average state (\fig{fig2}e). We specifically observe that the $12$ largest eigenvalues are significantly above the noise floor, a number that matches the number of existing optical modes in the area covered by the target. Several of these eigenvalues are close to be degenerate, as explained by symmetries in the intensity distributions generated by the eigenstates.

The generality of our formalism allows us to identify optimal states not only in the case of a single bead but also for more complicated targets. As an example, we study the case of a target composed of a cluster of six beads (\fig{fig2}g-l). In this case, the eigenstates that significantly interact with the target (\fig{fig2}h-j, see also~\cite{supplementary} Section~S5.2) are all characterized by complicated intensity distributions that cannot be easily predicted without the knowledge of the discrimination operator. These results illustrate the fact that optimal states do not necessarily focus light everywhere onto extended targets, but instead provide one with the unique optimal solutions that generally minimize the Helstrom bound, taking into account absorption as well as all single and multiple scattering effects (including, e.g., strong coupling effects occurring between a light field and the dipoles induced within a strongly scattering target). However, for such an extended target, the eigenvalue distribution appears as a continuum (\fig{fig2}g), which prevents us from easily relating the number of large eigenvalues to the spatial extent of the target.

\begin{figure}[t]
	\begin{center}
		\includegraphics[width=\linewidth]{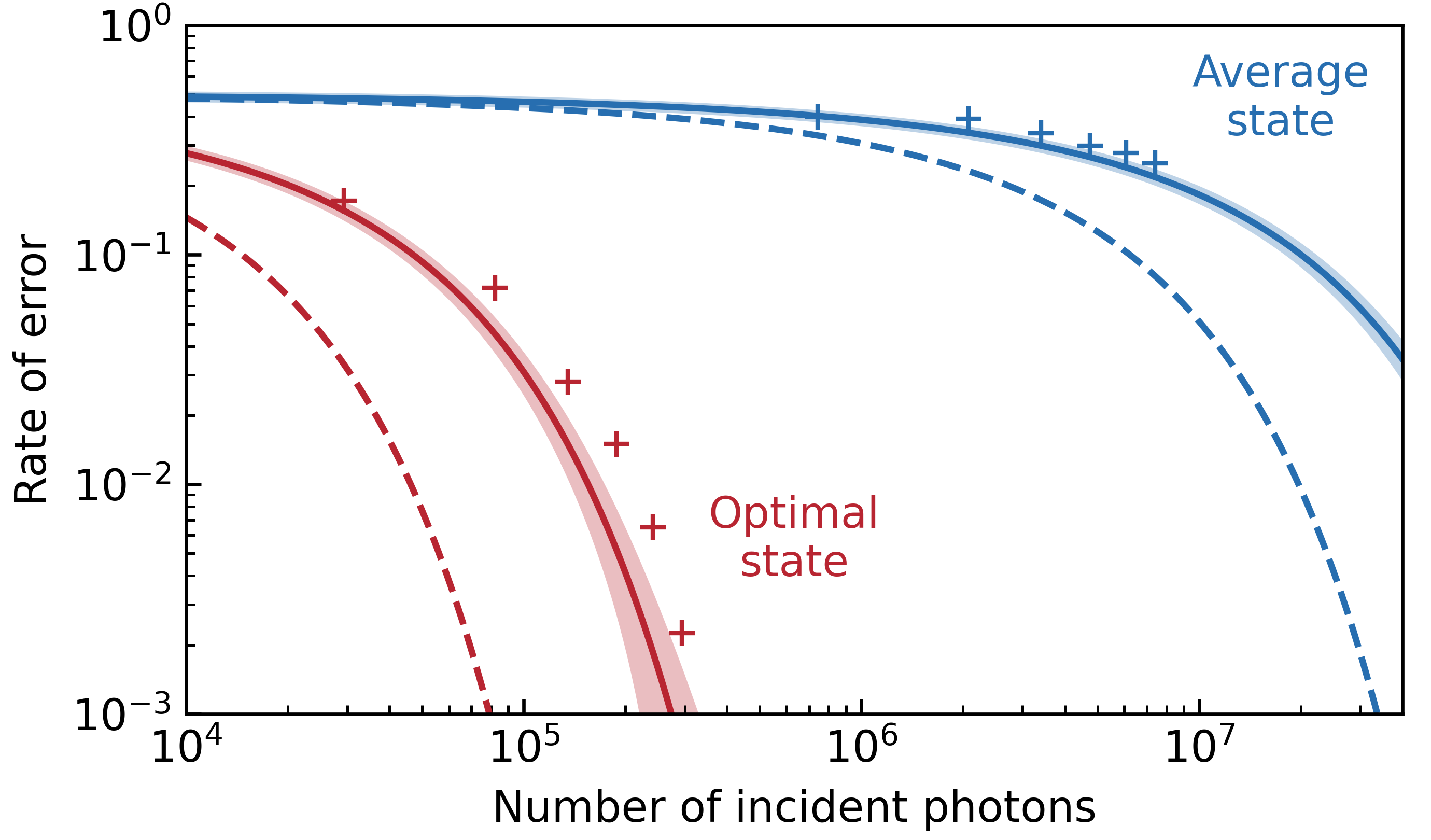}
	\end{center}
	\caption{Rate of error as a function of the number of incident photons for the average state (blue) and the optimal state (red). In both cases, the experimentally-observed error rates (data points) are compared to the theoretical values associated with our homodyne setup (solid lines) and to the Helstrom bound (dashed lines). Shaded areas represent $95.4$\% confidence intervals, taking into account only the statistical error caused by the finite number of measurements ($N_{\mathrm{rep}}=4000$). Note that, while we illuminate the system with up to $7.4\times10^6$ photons, many photons are scattered out of the field of view by the diffusers. As a consequence, we only detect up to $140$ photons over the area covered by the camera sensor.}
	\label{fig3}
\end{figure}

While we introduced the discrimination operator to minimize the Helstrom bound, which can be reached only with an optimal detection scheme, it is also the relevant operator to find optimal incident fields in the case of Gaussian receivers, which are sub-optimal but widely used due to their simplicity~\cite{weedbrook_gaussian_2012}. Indeed, a simple homodyne scheme is optimal among all available Gaussian receivers~\cite{takeoka_discrimination_2008}, whereas implementing an optimal detection scheme---such as a Dolinar receiver~\cite{cook_optical_2007}---typically requires low-noise time-resolved photon detection along with an excellent interferometric stability. With the homodyne detection scheme implemented in our experiment (which is shot-noise limited), measured field quadratures follow a Gaussian distribution of variance $\sigma^2=1/2$ (see~\cite{supplementary}, Section~S2.1). For two hypotheses with equal \textit{a priori} probabilities ($\pi_1=\pi_2=0.5$), the theoretical probability of error associated with this Gaussian receiver is then expressed by (see~\cite{supplementary}, Section~S2.2),
\begin{equation}
P_\mathrm{G} = \frac{1}{2} \erfc \left( \sqrt{ \frac{n d_{12}^2}{8 \sigma^2}} \; \right) .
\label{eq:error_homodyne}
\end{equation}
This expression shows that accessing the discrimination operator allows one to minimize not only the Helstrom bound but also the probability of error associated with homodyne detection schemes, as they are both governed by the statistical distance $d_{12}$. This is experimentally demonstrated by performing measurements in low-light conditions, using a variable attenuator to gradually change the number of incident photons. The presence of the bead located in between diffusers is tested by illuminating the system with either the optimal incident state ($d_{12}^2=\Lambda_1$) or with the average state ($d_{12}^2=\bar{\Lambda}$). Measured data are then processed using the likelihood-ratio test (see~\cite{supplementary}, Section~S2.3), which is theoretically optimal~\cite{trees_detection_2013}. The resulting rate of error observed over $N_{\mathrm{rep}}=4000$ measurements is shown in \fig{fig3} as a function of the number $n$ of incident photons. The measured rate of error, which exponentially decreases with $n$, is characterized by a decay constant that is larger by more than two orders of magnitude for the optimal state (red points) as compared to the average one (blue points). Consequently, with the optimal state, far fewer photons are needed to accurately detect the presence of the target. Measured rates of error are only slightly higher than theoretical values predicted by \eq{eq:error_homodyne} (solid lines), a difference that is due to a slightly sub-optimal decision criterion---the likelihood-ratio test requires unbiased estimates of the field expectation values that are difficult to obtain in low-light conditions. Finally, the Helstrom bound calculated using \eq{eq:error_helstrom} is similarly reduced when the optimal state is used instead of the average one (dashed lines), although with a lower overall error rate.

Whereas we measured only sub-parts of the full $S$-matrix in our experiments---as usually done in optics~\cite{popoff_measuring_2010,yu_measuring_2013}---it is also instructive to discuss the ideal case of unitary scattering matrices ($S_i^{-1}=S_i^\dagger$). This is achieved for systems without gain or loss (e.g. multimode fibers~\cite{ploschner_seeing_2015,matthes_learning_2021}), if one has access to all existing incident and outgoing optical modes. The eigenstates of $D_{12}$ are then solutions of a generalized linear eigenvalue problem, in the same way as scattering invariant modes~\cite{pai_scattering_2021}. In this unitary limit, the optimal state along with all other eigenstates of $D_{12}$ thus share a remarkable property: they all produce outgoing fields that are independent of which of the two scattering systems they interact with, except for a global phase factor which affects all outgoing modes and which contains all available information (see~\cite{supplementary}, Section~1.2). Nevertheless, as could be expected from the sub-unitarity of the measured transmission matrices, this property is not observed in our experiment (see Fig~1c,d).

% =====================================================================================
% conclusion
% =====================================================================================

To summarize, we demonstrated how to spatially modulate light fields in order to optimally discriminate between different configurations of a complex scattering system. To this end, we introduced the discrimination operator $D_{12}$, which quantifies the amount of information produced by any perturbation of a discrete observable. We experimentally showed how to use this operator for generating light states that are optimally tailored to detect a target located inside a disordered medium even in low-light conditions. These results open up new perspectives to improve the performances of nanophotonic sensing devices based, e.g., on photonic crystal waveguides~\cite{bag_towards_2020}, metasurfaces \cite{wolterink_localizing_2021,buijs_programming_2021}, or cavities~\cite{del_hougne_deeply_2021}. The orthonormal basis formed by the eigenstates of $D_{12}$ is also well-suited to analyze experiments based on time-reversed adapted perturbation~\cite{fink_time-reversed_2000,zhou_focusing_2014,ma_time-reversed_2014,ruan_focusing_2017}, which in turn suggests interesting experimental approaches to generate optimal states. Moreover, by associating scattering matrix measurements with advanced optimization procedures~\cite{bouchet_optimizing_2021}, optimal states could potentially be identified for decision problems involving more than two configurations as well as to composite hypotheses problems~\cite{helstrom_quantum_1976,trees_detection_2013}. Interestingly, our results might also find applications in cryptography, notably to identify physical unclonable keys that are the most difficult to reproduce~\cite{pappu_physical_2002,uppu_asymmetric_2019}. Finally, the formalism developed in our work suggests a new path to study complex scattering systems using quantum illumination~\cite{pirandola_advances_2018}. In this perspective, the use of squeezed states of light emerges as a promising approach~\cite{andersen_30_2016,chesi_squeezing-enhanced_2018}.

% =====================================================================================
% acknowledgments
% =====================================================================================

\begin{acknowledgments}
	\paragraph*{Acknowledgements.}
	The authors thank Ir\`ene Wang for insightful discussions, and Philippe Moreau for technical support. This work was supported by the European Research Council (ERC) within the H2020 program (grant 681514-COHERENCE), by the Nederlandse Organisatie voor Wetenschappelijk Onderzoek NWO (Vici 68047618) and by the Austrian Science Fund (FWF) under project number P32300 (WAVELAND).
\end{acknowledgments}

% =====================================================================================
% references
% =====================================================================================

\bibliographystyle{apsrev4-2}
%\bibliography{references/references}

%

	% =====================================================================================
	% supplementary
	% =====================================================================================
	
	\onecolumngrid
	\pagebreak
	\beginsupplement
	\begin{center}
		\textbf{\large Optimal control of coherent light scattering for binary decision problems\\ \bigskip Supplementary information}
		
		\bigskip
		Dorian Bouchet,$^1$ Lukas M.\ Rachbauer,$^2$ Stefan Rotter,$^2$ Allard P.\ Mosk,$^3$ and Emmanuel Bossy$^1$\\ \vspace{0.15cm}
		\textit{\small $^\mathit{1}$Université Grenoble Alpes, CNRS, LIPhy, 38000 Grenoble, France}\\ \smallskip
		\textit{\small $^\mathit{2}$Institute for Theoretical Physics, Vienna University of Technology (TU Wien), 1040 Vienna, Austria}\\ \smallskip
		\textit{\small $^\mathit{3}$Nanophotonics, Debye Institute for Nanomaterials Science and Center for Extreme Matter and Emergent Phenomena, Utrecht University, P.O. Box 80000, 3508 TA Utrecht, Netherlands}
	\end{center}
	\vspace{0.3cm}

\section{Optimal incident state}

\subsection{General case}

We define the incident field state $| E^\ins \ket$ in the Hilbert space of all incident spatial modes. This state is characterized by the coefficients $\lbrace E^\ins_1,\dots,E^\ins_M \rbrace$, defined as the expectation values of the field operator in $M$ incoming spatial modes. After interacting with a scattering system, such an incident state produces an outgoing field state $|E_i^\out \ket$, defined in the Hilbert space of all outgoing spatial modes, and where $i$ denotes the configuration of the scattering system interacting with the field ($i=1$ if $H_1$ is true, and $i=2$ if $H_2$ is true). Outgoing field states are characterized by the coefficients $\lbrace E^\out_{i,1},\dots,E^\out_{i,N} \rbrace$, defined as the expectation values of the field operator in $N$ outgoing spatial modes. As a convention, we express the fields in units of $\sqrt{(2 \hbar \omega)/(\epsilon_0 c_0 \Delta t A)}$, where $\hbar$ is the reduced Planck constant, $\omega$ is the angular frequency of the field, $\epsilon_0$ is the vacuum permittivity, $c_0$ is the speed of light in vacuum, $\Delta t$ is the integration time and $A$ is the effective mode area~\cite{loudon_quantum_2000_2}. In this way, the average numbers of photons in the incident and outgoing states are expressed by $\bra E^\ins|E^\ins \ket $ and $\bra E^\out|E^\out \ket $, respectively. Using a scattering matrix formalism, incident and outgoing field states are related by the following expression:
\begin{equation}
|E_i^\out \ket = S_i | E^\ins \ket ,
\end{equation}
where $S_i$ is the scattering matrix associated with the $i$-th hypothesis. In order to separately study the role of the total number of photons $n=\bra E^\ins|E^\ins \ket$ in the incident state and that of its spatial distribution, we define the normalized incident state $|\mathcal{E}^\ins\ket=n^{-1/2} | E^\ins \ket$ so that $\bra \mathcal{E}^\ins |\mathcal{E}^\ins\ket=1$. Writing $E_{i,k}^\out $ as a projection of $|E_i^\out \ket$ on the state $|k\ket$ associated with the $k$-th spatial mode, we obtain
\begin{equation}
E_{i,k}^\out = \sqrt{n} \, \bra k | S_i | \mathcal{E}^\ins \ket .
\label{eq_proj}
\end{equation} 
The statistical distance $d_{12}$ is defined as follows:
\begin{equation}
d_{12}^2 = \frac{1}{n} \sum_{k=1}^N \left \vert E_{2,k}^\out - E_{1,k}^\out \right \vert ^2 .
\label{def_distance}
\end{equation}
Inserting \eq{eq_proj} into \eq{def_distance} yields
\begin{equation}
d_{12}^2= \sum_{k=1}^N \left \vert \bra k | S_2 - S_1 | \mathcal{E}^\ins \ket \right \vert ^2 .
\end{equation}
This expression can be expanded into 
\begin{equation}
d_{12}^2= \sum_{k=1}^N \bra \mathcal{E}^\ins | (S_2-S_1)^\dagger | k \ket \bra k | S_2-S_1 | \mathcal{E}^\ins \ket .
\end{equation}
Using the completeness relation $\sum_k | k \ket \bra k |=I_N$ where $I_N$ is the $N$-dimensional identity matrix, we finally obtain
\begin{equation}
d_{12}^2 = \bra \mathcal{E}^\ins | D_{12} |\mathcal{E}^\ins \ket ,
\end{equation}
where we introduced the discrimination operator 
\begin{equation}
D_{12}= \left(S_2-S_1\right)^\dagger ( S_2-S_1) .
\end{equation}
This operator, which is Hermitian by construction, quantifies the statistical distance between two scattering matrices. Its largest eigenvalue is equal to the maximum value of $d_{12}^2$ that can be reached by shaping the incident field state in its spatial degrees of freedom, and the eigenvector associated with this eigenvalue gives the spatial distribution of this optimal incident field state.

\subsection{Unitary limit}

For two scattering matrices that are unitary ($S_i^\dagger = S_i^{-1}$), the operator $D_{12}$ is expressed by 
\begin{equation}
D_{12}= 2 I_N - 2 \re (S_2^\dagger S_1 ) ,
\end{equation}
where $\re(S_2^\dagger S_1)$ denotes the Hermitian part of $S_2^\dagger S_1$. In this case, the eigenstates of $D_{12}$ satisfy the eigenvalue equation:
\begin{equation}
(S_2^\dagger S_1 + S_1^\dagger S_2 ) | \mathcal{E}^\ins \ket = (2-\Lambda) |\mathcal{E}^\ins \ket ,
\end{equation}
where $\Lambda \in \mathbb{R}$ denotes an eigenvalue of $D_{12}$. This equation can be identified as the eigenvalue equation for scattering invariant modes \cite{pai_scattering_2021_2}. Since both $S_1$ and $S_2$ are unitary, the eigenstates of $D_{12}$ satisfy the following generalized eigenvalue equation:
\begin{equation}
S_2 |\mathcal{E}^\ins \ket = e^{i \theta} S_1 |\mathcal{E}^\ins \ket ,
\end{equation}
where $\theta \in \mathbb{R}$. Thus, when eigenstates of $D_{12}$ propagate into each scattering system, the resulting outgoing field states satisfy $|E^\out_2\ket = e^{i \theta} |E^\out_1\ket$, which shows that both outgoing fields are identical except for a phase change of $\theta$. This phase change is related to the eigenvalue $\Lambda$ by the following relation:
\begin{equation}
\Lambda = 2 (1-\cos \theta) .
\end{equation}
The statistical distance $d_{12}$ is minimum when $\theta=2 \pi m $, $m \in \mathbb{Z}$; in such case, $d_{12}^2=0$ and the phase of the outgoing state does not depend on the scattering system the wave propagates in. In contrast, the statistical distance $d_{12}$ is maximum when $\theta=\pi + 2 \pi m$, $m \in \mathbb{Z}$; in such case, $d_{12}^2=4$ due to a phase difference of $\pi$ in the outgoing state when the scattering system is changed.

% =====================================================================================
% section
% =====================================================================================

\section{Rate of error for the homodyne scheme}

\subsection{Noise statistics}

In the experiment, we implemented a homodyne scheme based on digital off-axis holography. In the shot-noise limit and with a strong reference beam ($|E^\rf_k|^2\gg|E^\out_{i,k}|^2$), the complex field retrieved by such homodyne scheme can be modeled by an $N$-dimensional complex random variable $Z$ such that~\cite{goodman_statistical_2015_2}
\begin{subequations}
	\begin{align}
	\re Z_{k} &\sim \mathcal{N} \left( \re E^{\out}_{1,k}, \sigma^2 \right) \; \mathrm{and} \; \im Z_{k} \sim \mathcal{N} \left( \im E^{\out}_{1,k}, \sigma^2 \right) \; \mathrm{if} \; H_1 \; \mathrm{is} \; \mathrm{true}  , \\
	\re Z_{k} &\sim \mathcal{N} \left( \re E^{\out}_{2,k}, \sigma^2 \right) \; \mathrm{and} \; \im Z_{k} \sim \mathcal{N} \left( \im E^{\out}_{2,k}, \sigma^2 \right) \; \mathrm{if} \; H_2 \; \mathrm{is} \; \mathrm{true} ,
	\end{align}
\end{subequations}
where $\sigma^2=1/2$. In practice, for measured data to follow these statistics, it is required to determine the value of the internal gain of the camera, which is approximately $8.8$ photo-electrons per digital count for our camera (Basler acA1300-200um). Multiplying measured images by this gain factor ensures that intensities are expressed in terms of numbers of photons---the sub-unitary quantum efficiency of the camera being then implicitly included in the definition of measured transmission matrices. 

\begin{figure}[ht]
	\begin{center}
		\includegraphics[width=0.97\linewidth]{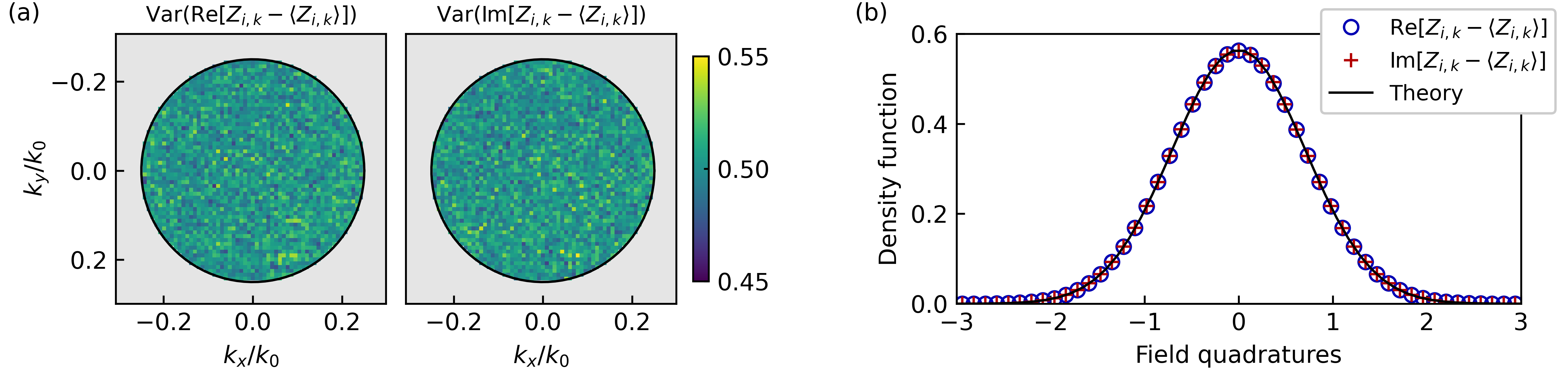}
	\end{center}
	\caption{(a)~Variance of the real part and the imaginary part of the field, experimentally measured with a homodyne scheme in an off-axis configuration. $k_0$ denotes the norm of the wavevector, while $k_x$ and $k_y$ denote its components along the $x$ and $y$ directions, respectively. (b)~Measured distributions of the field quadratures, after subtraction of the mean fields. The theoretical distribution is a centered normal distribution of variance $\sigma^2=1/2$. All figures are obtained from $N_{\mathrm{rep}}=4000$~measured fields composed of $N=2617$~spatial modes. These fields were acquired using the optimal incident state and $n=2.9 \times 10^5$ incident photons.}
	\label{fig:noise}
\end{figure}

To demonstrate that the data measured with our setup follow these statistics, we considered a data set composed of $4000$ fields measured in low-light conditions. Among them, $2000$ fields were measured in the presence of the target, and $2000$ fields were measured without the target. After subtracting the mean fields associated with the two different scattering systems, we calculated the variance of field quadratures for each outgoing mode (\fig{fig:noise}a). It clearly appears that the measured variance is uniform, and that its value is in excellent agreement with the theoretical value $\sigma^2=1/2$. Taking into account all modes within the field of view, we also verified that the measured distribution of the field quadratures is a centered normal distribution (\fig{fig:noise}b). Results presented here were obtained with the optimal incident state, a target composed of a single bead, and $n=2.9 \times 10^5$ incident photons. Nevertheless, similar results were obtained for all data presented in Fig.~3 of the manuscript, that were all acquired in low-light conditions. 

\subsection{Theoretical probability of error}

The minimum probability of error that can be achieved when choosing a hypothesis from measured data is expressed by~\cite{trees_detection_2013_2}
\begin{equation}
P_\mathrm{G} = \frac{\pi_1}{2} \erfc \left[ \sqrt{ \frac{n d_{12}^2}{8 \sigma^2}} + \ln\left(\frac{\pi_1}{\pi_2}\right) \sqrt{\frac{\sigma^2}{2n d_{12}^2}} \; \right] +\frac{\pi_2}{2} \erfc \left[ \sqrt{ \frac{n d_{12}^2}{8 \sigma^2}} - \ln\left(\frac{\pi_1}{\pi_2}\right) \sqrt{\frac{\sigma^2}{2n d_{12}^2}} \; \right] ,
\label{proba_error}
\end{equation}
where $\pi_1$ and $\pi_2$ are the \textit{a priori} probabilities associated with each hypothesis. Assuming that $\pi_1=\pi_2=0.5$, we obtain
\begin{equation}
P_\mathrm{G} = \frac{1}{2} \erfc \left( \sqrt{ \frac{n d_{12}^2}{8 \sigma^2}} \; \right) .
\label{eq:theoretical_rate}
\end{equation}
When estimating the probability of error $P_\mathrm{G}$ from a finite number of trials $N_{\mathrm{rep}}$, the number of errors that are observed follows a binomial distribution. The variance of the estimate $\hat{P}_\mathrm{G}$ is then given by 
\begin{equation}
\Ve \left(\hat{P}_\mathrm{G}\right)=\frac{P_\mathrm{G}(1-P_\mathrm{G})}{N_{\mathrm{rep}}} .
\end{equation}
For $N_{\mathrm{rep}}$ sufficiently large, the probability distribution of $\hat{P}_\mathrm{G}$ approaches a normal distribution. This property allows us to define the following $95,4$\% confidence interval:
\begin{equation}
\left[ P_\mathrm{G}- 2 \sqrt{\frac{P_\mathrm{G}(1-P_\mathrm{G})}{N_{\mathrm{rep}}}} ; P_\mathrm{G}+ 2 \sqrt{\frac{P_\mathrm{G}(1-P_\mathrm{G})}{N_{\mathrm{rep}}}} \right] .
\end{equation}

\subsection{Measured rate of error}

The rate of error observed in the experiment is obtained by processing noisy data $Z$ using the likelihood-ratio test, which theoretically reaches the bound expressed by \eq{proba_error}. The decision criterion is given by
\begin{subequations}
	\label{eq:decision}
	\begin{align}
	\ln l(Z) &< \ln\left(\frac{\pi_1}{\pi_2}\right) \rightarrow \mathrm{choose}\; H_1 , \\
	\ln l(Z) &> \ln\left(\frac{\pi_1}{\pi_2}\right) \rightarrow \mathrm{choose}\; H_2 , \\
	\ln l(Z) &= \ln\left(\frac{\pi_1}{\pi_2}\right) \rightarrow \mathrm{choose \; either}\; H_1 \; \mathrm{or} \; H_2 ,
	\end{align}
\end{subequations}
where $\ln l(Z)$ is the log-likelihood ratio expressed by~\cite{trees_detection_2013_2}
\begin{equation}
\ln l(Z) = \re \left[ \sum_{k=1}^N \frac{ (E_{2,k}^\out - E_{1,k}^\out)^* Z_{k} }{\sigma^2} \right] + \sum_{k=1}^N \frac{\vert E_{1,k}^\out \vert^2-|E_{2,k}^\out|^2}{2\sigma^2} .
\label{eq:likelihood}
\end{equation}
Calculating the log-likelihood ratio requires the knowledge of $E^\out_{1,k}$ and $E^\out_{2,k}$, which are the expectation values of the field under each hypothesis and for each outgoing spatial mode. These fields can be expressed as follows:
\begin{subequations}
	\begin{align}
	E_{1,k}^\out &= \frac{E^\out_{\mathrm{s},k} - E^\out_{\mathrm{d},k}}{2 \pi_1} , \\[0.1in]
	E_{2,k}^\out &= \frac{E^\out_{\mathrm{s},k} + E^\out_{\mathrm{d},k}}{2 \pi_2} , 
	\end{align}
\end{subequations}
where we introduced $E^\out_{\mathrm{s},k} = \pi_1 E_{1,k}^\out+\pi_2 E_{2,k}^\out$ and $E^\out_{\mathrm{d},k}=\pi_2 E_{2,k}^\out- \pi_1 E_{1,k}^\out$. There are different possible strategies to assess $E^\out_{\mathrm{s},k}$ and $E^\out_{\mathrm{d},k}$. A straightforward strategy entails estimating both $E^\out_{\mathrm{s},k}$ and $E^\out_{\mathrm{d},k}$ from measurements performed with a large number of incident photons. Here, we opted for a different strategy, in which $E^\out_{\mathrm{d},k}$ is estimated from measurements performed with a large number of incident photons, but with $E^\out_{\mathrm{s},k}$ being directly assessed from the data measured in low-light conditions by averaging them over noise fluctuations ($E^\out_{\mathrm{s},k} \simeq\langle Z_{k} \rangle$). This strategy, which is relevant only when $N_{\mathrm{rep}}$ is sufficiently large, allows to reduce biases that are observed when processing experimental data, including those due to unwanted reflections of the reference field by the camera sensor, to wavefront distortions generated by the optical elements used to control the incident photon flux, and to power fluctuations of the incident laser beam. Nevertheless, biases can also appear in this case due to the finite number of measurements used to estimate $E^\out_{\mathrm{s},k}$, resulting in an observed rate of error that is slightly higher than the theoretically-predicted one (see Fig.~3 of the manuscript).

% =====================================================================================
% section
% =====================================================================================

\section{Experimental implementation}

\subsection{Optical setup}

\begin{figure}[ht]
	\begin{center}
		\includegraphics[width=\linewidth]{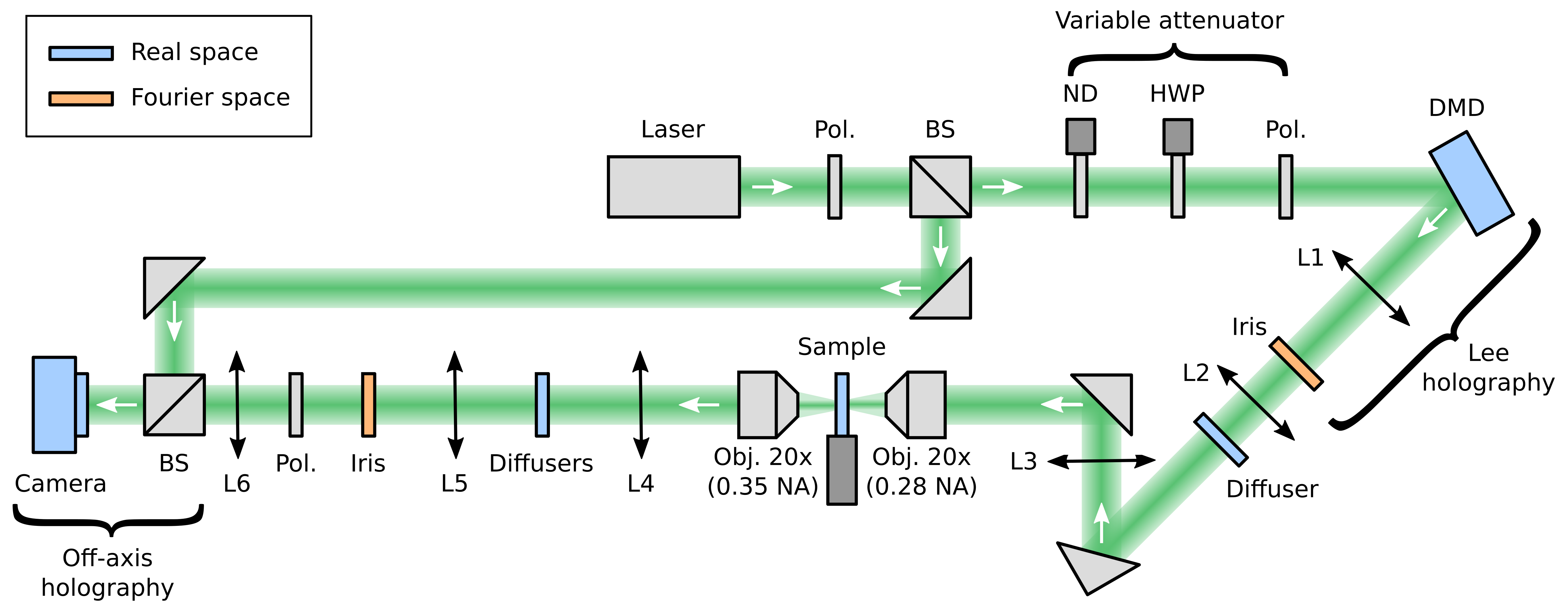}
	\end{center}
	\caption{Schematic of the optical setup used to acquire transmission matrices and perform measurements in low-light conditions. The sample is mounted on a motorized translation stage. The incident field is modulated with a digital micromirror device (DMD) using Lee holography, and the outgoing field is measured by a camera using off-axis holography. Diffusers are placed in intermediate image planes before and after the sample (this configuration optically imitates a situation in which the sample is located within a disordered material). The incident photon flux is controlled with a variable attenuator. Pol, linear polarizer; BS, beamsplitter; ND, neutral density filters; HWP: half wave-plate; Obj, microscope objective; NA, numerical aperture; L1 to L6, lenses.}
	\label{fig:setup}
\end{figure}

The optical setup used to acquire transmission matrices and perform measurements in low-light conditions is represented in \fig{fig:setup}. The sample under study is composed of polystyrene beads dispersed on a glass coverslip. To prepare this sample, we used a commercial solution of polystyrene beads (Polysciences Polybead, diameter 3\,\textmu m $\pm150$\,nm) that we diluted into 99\% isopropyl alcohol. We then deposed it onto a clean glass coverslip and let it dry; this results in a sample with a very low density of beads (approximately 1 bead inside a 100\,\textmu m$\times$100\,\textmu m area). Using this procedure, some of these beads are isolated, while others are aggregated into clusters. This allows us to study both the case of one isolated bead, and the case of a cluster of beads.

The light source used in our experiments is a continuous-wave solid-state laser (Cobolt 08-DPL) emitting at $\lambda=532$\,nm. The laser light is coupled to a single-mode polarization-maintaining fiber and out-coupled using a collimator (Schäfter+Kirchhoff 60FC-L-4-M75-01). A linear polarizer is used to ensure that the light is linearly polarized. The beam is separated into a signal path and a reference path using a 90:10 (transmission:reflection) beamsplitter. 

In the signal path, the light beam passes through a variable attenuator composed of a neutral density filter of fractional transmittance $\mathcal{T}_\mathrm{nd}=10^{-3.6}$ mounted on a motorized flip mount (Thorlabs MFF101/M), a half wave-plate mounted on a motorized rotation mount (Thorlabs PRM1/MZ8) and a linear polarizer. The light beam is reflected and modulated with a DMD (Vialux superspeed V-7001) using Lee holography~\cite{lee_binary_1974_2}, at a rate of 1400\,Hz. Light passes through a 4f system composed of a 200\,mm lens (L1) and 30\,mm lens (L2). An iris located in the focal plane in-between the two lenses selects the first diffraction order of the grating displayed by the DMD. 

A first scattering layer composed of one ground glass diffuser (Thorlabs DG20-1500, 1500 grits) can be placed in the focal plane of L2. This plane is optically conjugated with the sample plane using a 200\,mm lens (L3) and a $\times$20 objective (Mitutoyo Plan Apo SL 20X/0.28). The sample is mounted on a motorized translation stage (PI M-122.2DD1). The sample plane is optically conjugated with an intermediate image plane, using a $\times$20 objective (Nikon CF Plan 20X/0.35 EPI SLWD) and a 100\,mm lens (L4). A second scattering layer composed of two consecutive ground glass diffusers (Thorlabs DG10-600, 600 grits) can be placed in this intermediate image plane. This configuration, with two scattering layers optically conjugated with the sample plane, resembles a situation in which a (moving) sample of interest is located within a (static) disordered material.

Light then passes through a 4f system composed of a 100\,mm lens (L5) and 200\,mm lens (L6). An iris located in the focal plane in-between the two lenses blocks the light scattered at high angles by the second scattering layer. After passing through a linear polarizer, a 90:10 (transmission:reflection) beamsplitter is used to recombine the reference path with the signal path. The resulting intensity pattern is measured using a complementary metal oxide semiconductor camera (Basler acA1300-200um) with an exposure time of 550\,\textmu s; both quadratures of the complex field are then reconstructed using digital off-axis holography~\cite{cuche_spatial_2000_2}. In order to reduce the influence of unwanted reflections of the reference beam by the camera sensor, all basic optical components have an antireflective coating, and a small angle is introduced between the beam and the normal to the camera. 

\subsection{Measured intensity patterns with plane-wave illumination}

\begin{figure}[ht]
	\begin{center}
		\includegraphics[width=\linewidth]{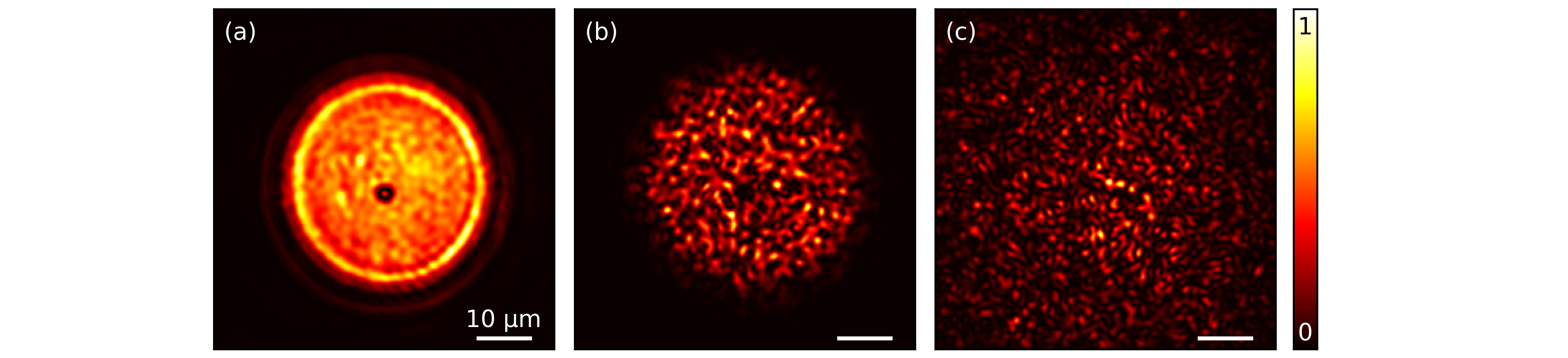}
	\end{center}
	\caption{Intensity distributions measured in the presence of the target for plane wave illumination (a)~in the absence of both scattering layers, (b)~in the presence of the first scattering layer only and (c)~in the presence of both scattering layers.}
	\label{fig:distributions_pw}
\end{figure}

In order to give a better insight of how complex the system is and how much the beam is spread at the bead location, we use a (clipped) plane wave of normal incidence to illuminate the target without any scattering layer (\fig{fig:distributions_pw}a), with only the first scattering layer located between the DMD and the sample (\fig{fig:distributions_pw}b), and with both scattering layers (\fig{fig:distributions_pw}c). This clearly shows that both scattering layers are complex, since the plane wave becomes a speckle after passing through the first scattering layer, and the speckle is fully different after passing through the second scattering layer. The beam spread induced by the first scattering layer (composed on one single diffuser) is relatively small; in this way, we can be sure that no more than one bead interacts with the incident field (a few other beads are also located on the same sample, but outside the field of view). In contrast, the beam spread induced by the second scattering later is considerably larger, due to the presence of two contiguous diffusers in-between the sample and the camera. 

\section{Acquisition procedure}

The acquisition procedure consists of three main steps:
\begin{enumerate}
	\itemsep0em 
	\item We measure transmission matrices with and without the target, with a large number of incident photons. This step allows us to access the discrimnation operator $D_{12}$, from which optimal and average incident states are calculated. 
	\item We generate the average field state as well as the first $15$ eigenstates, with and without the target, and with a large number of incident photons. This step allows us to ensure that the field states predicted from the knowledge of $D_{12}$ can be faithfully generated with our experimental setup. We also perform the same measurements in the absence of the second scattering layer, in order to measure the field distribution in the target plane.
	\item We drastically reduce the number of incident photons and we perform a large number of measurements, with and without the target, using the optimal incident state as well as the average one. This step allows us to experimentally quantify the rate of error achievable with each field state as a function of the number of incident photons. 
\end{enumerate}

All three steps are performed by running the setup at $1400$\,Hz, resulting in an effective acquisition rate of $700$\,Hz. Indeed, for each field that we want to measure, we also acquire a phase-reference field that is generated using a plane wave of normal incidence. This procedure allows us to monitor how the global phase changes over time due to temperature, mechanical and laser wavelength drifts. Global phase drifts are calculated from the complex inner product of all phase-reference fields with the phase-reference field measured at the beginning of the acquisition. Spline interpolations are then used to estimate and correct for the global phase drift at any time during the acquisition. This procedure yields an inter-frame phase error of approximately $0.01$\,rad.

\subsection{Acquisition of transmission matrices}

We measure two (sub-unitary) transmission matrices $S_1$ and $S_2$, relating incident field states to transmitted ones~\cite{popoff_measuring_2010_2}. The matrix $S_1$ is measured without the target in the field of view, and the matrix $S_2$ is measured after translating the target inside the field of view. In our experiment, these two transmission matrices are measured in 5.3\,s. Note that this time is limited by the acquisition rate of the camera, and could thus be reduced by using a high-speed camera (one could also use a fast single-channel detector since the number of outgoing spatial modes can be taken as low as $N = 1$). Measurements of $S_1$ and $S_2$ are performed with no density filter in the signal path, and thus with a large number of incident photons. To illuminate the scattering system, we vary the incidence angle of a plane wave that is clipped to a diameter of $40$\,\textmu m in the sample plane. More precisely, we sample the incident field in Fourier space using a triangular lattice~\cite{pai_optical_2020_2}, covering a numerical aperture of $\mathrm{NA}=0.25$ with $M=1735$ different incidence angles. For each angle, we record the transmitted field using digital off-axis holography. This method relies on a reference beam that is tilted by an angle with respect to the signal beam. With this approach, the complex field can be directly accessed in Fourier space by selecting the first-order component. We therefore sample the transmitted field in Fourier space, with the square lattice defined by the pixels of the camera, covering a numerical aperture of $\mathrm{NA}=0.25$ with $N=2617$ sampling points. Transmission matrices are therefore measured column by column and, as a result, we obtain $2617\times1735$ transmission matrices. We normalize these matrices by dividing them by $\sqrt{n_0}$, where $n_0$ is the number of incident photons associated with each plane wave used to construct the transmission matrices. This number was measured beforehand with a powermeter ($ n_0 = 2.2 \times 10^{11}$ photons).

\subsection{Verification of predicted field states}

From the knowledge of the two transmission matrices $S_1$ and $S_2$, the discrimination operator $D_{12}=(S_2-S_1)^\dagger (S_2-S_1)$ is easily calculated (processing time of 215\,s). We specifically study the average state, defined as an equally-weighted linear superposition of all eigenstates of $D_{12}$, as well as the first $15$ eigenstates of $D_{12}$ (including the optimal state, which is the first eigenstate of $D_{12}$). We experimentally generate these states and, for each of them, we perform $50$ measurements with and without the target, first in the presence of both scattering layers (in order to compare experimentally-generated light states to predicted ones) and then in the absence of the scattering layer located after the sample (in order to directly measure the field distribution in the target plane).

Averaging over measurements performed in the presence of both scattering layers allows us to compare experimentally-generated light states $|E_i^{\mathrm{out,meas}}\ket$ to the predicted ones $|E_i^{\mathrm{out,pred}}\ket = \sqrt{n_0} S_i |\mathcal{E}^\ins\ket $. This comparison is achieved by calculating the complex correlation coefficient $\mathscr{C}_i$ and the squared norm ratio $R_i$, that are respectively expressed by
\begin{subequations}
	\begin{gather}
	\mathscr{C}_i = \frac{\bra E_i^{\mathrm{out,pred}} |E_i^{\mathrm{out,meas}} \ket }{\Vert E_i^{\mathrm{out,pred}} \Vert . \Vert E_i^{\mathrm{out,meas}} \Vert } , \\[0.1in]
	R_i = \frac{\Vert E_i^{\mathrm{out,meas}} \Vert^2}{\Vert E_i^{\mathrm{out,pred}} \Vert^2 } .
	\end{gather}
\end{subequations}
Results for the average state and the optimal state are presented in \tab{tab:fidelity}. 

\begin{table}[ht]
	
	\caption{\label{tab:fidelity}Fidelity of the experimental generation of light states.}
	
	\begin{ruledtabular}
		\begin{tabular}{C{0.04\textwidth}|C{0.15\textwidth}C{0.15\textwidth}|C{0.15\textwidth}C{0.15\textwidth}}
			& \multicolumn{2}{c|}{Single bead} & \multicolumn{2}{c}{Six beads}\\ \hline 
			& Average state & Optimal state & Average state & Optimal state \\
			$|\mathscr{C}_{1}|$ & $0.97$ & $0.96$ & $0.97$ & $0.96$ \\
			$|\mathscr{C}_{2}|$ & $0.97$ & $0.95$ & $0.97$ & $0.93$ \\
			$R_{1}$ &$1.3 \times 10^{-1}$ & $5.2 \times 10^{-3}$ & $1.3 \times 10^{-1}$ & $7.2 \times 10^{-3}$ \\
			$R_{2}$ &$1.3 \times 10^{-1}$ & $5.4 \times 10^{-3}$ & $1.3 \times 10^{-1}$ & $7.4 \times 10^{-3}$ \\
			$\eta_d$ & $0.98$ & $0.95$ & $0.97$ & $0.91$ \\
		\end{tabular}
	\end{ruledtabular}
\end{table}

It clearly appears that the shape of outgoing states is faithfully generated by the DMD, with values of $|\mathscr{C}_i|$ between $0.9$ and $1$. However, the total intensity experimentally measured is significantly lower than the predicted one, with $R_{i} \simeq 10^{-1}$ for average states and $R_{i} \simeq 5\times 10^{-3}$ for optimal ones. This is explained by the low photon efficiency of techniques based on Lee holography to generate amplitude-and-phase modulated fields~\cite{lee_binary_1974_2,mirhosseini_rapid_2013_2}: for shaped waves, the actual number of incident photons is equal to $\mathcal{T}_{\mathrm{mod}} \, n_0$, where $\mathcal{T}_{\mathrm{mod}}$ is the fractional transmittance of the modulation technique. Indeed, while phase variations are encoded in the period of a binary grating, amplitude variations are encoded by deflecting photons out of the optical path. The number of incident photons is thus larger for plane waves than for shaped waves. Furthermore, the intensity distribution of optimal states is more spatially localized than those of average states, resulting in a lower photon efficiency of the modulation technique. In practice, the squared norm ratio $R_{1}$ can be taken as an estimate of the fractional transmittance $\mathcal{T}_{\mathrm{mod}}$ (the choice of $R_{1}$ over $R_{2}$ is made based on the consideration that values measured in the absence of the target are free of possible positioning errors of the translation stage). Finally, we calculate the ratio $\eta_d$ between measured and predicted values for $d_{12}^2$. We observe that values of $\eta_d$ are very close to unity, demonstrating that $d_{12}^2$ is faithfully estimated from transmission matrix measurements. 

\subsection{Measurements in low-light conditions}

We finally perform many measurements in low-light conditions. To this end, we place a neutral density filter of fractional transmittance $\mathcal{T}_{\mathrm{nd}}=10^{-3.6}$ in the signal path, and we use the variable attenuator to reduce even more the number of incident photons, with a fractional transmittance $\mathcal{T}_{\mathrm{va}}$ evenly varied $6$ times between $0.1$ and $1$. The incident number of photons is thus expressed by $n=\mathcal{T}_{\mathrm{nd}} \mathcal{T}_{\mathrm{va}} \mathcal{T}_{\mathrm{mod}} \, n_0$. In this way, we vary the number of incident photons from 740,000 to 7,400,000 when the average wave is used, and from 29,000 to 290,000 when the optimal wave is used (the difference between values obtained for the average wave and for the optimal wave is due to different values of $\mathcal{T}_{\mathrm{mod}}$). Since many photons are scattered out of the field of view by the diffusers, only a few photons are actually detected by the camera: the number of detected photons ranges from 14 to 140 when using the average wave, and from 2 to 20 when using the optimal wave. We thus detect on average 1 photon for 53,000 incident photons with the average wave, and 1 photon for 15,000 incident photons with the optimal wave. This demonstrates that the optimal state not only leads to an increased interaction between the light and the object, but also more efficiently redirects the light towards the observer, resulting in a larger ratio between detected and incident photons.

For each value of $n$, we successively generate the average state and the optimal state and, for each state, we perform $N_{\mathrm{rep}}=4000$ measurements. As we assume that the \textit{a priori} probabilities for each hypothesis are given by $\pi_1=\pi_2=0.5$, this results in the acquisition of $2000$ measurements in the presence of the target and $2000$ measurements in the absence of the target. In our experiment, this large data set ($2\times6\times4000$ measurements) is measured in 104\,s. We then calculate the log-likelihood ratio for each measured field with \eq{eq:likelihood}, and we deduce the experimental rate of error based on the decision criterion expressed by \eq{eq:decision} (processing time of 174\,s). This is finally compared to the theoretical rate of error expressed by \eq{eq:theoretical_rate}, where $\sigma^2=0.5$ and $d_{12}^2=\eta_d \langle \mathcal{E}^\ins | D_{12} | \mathcal{E}^\ins \ket$.

\section{Intensity distribution of eigenstates in the target plane}

\subsection{Small target}

By performing measurements in the absence of the second scattering layer, we can directly access the intensity distribution of the eigensates of $D_{12}$ in the target plane. The number of significant eigenstates is theoretically determined by the number of modes in the area $A_\mathrm{t}$ covered by the target, which can be approximated by $N_{\mathrm{t}} \simeq 2 \pi A_\mathrm{t} \mathrm{NA}^2 / \lambda^2$~\cite{mosk_controlling_2012_2}. Using this expression, we obtain a number of modes of the order of $12$ for a target composed of a single bead, in agreement with the observed number of eigenvalues that are significantly above the noise floor (see Fig.~2a of the manuscript). It is interesting to study the spatial distribution of the intensity associated with these largest eigenvalues, as such light fields significantly interact with the target.

\begin{figure}[b]
	\begin{center}
		\includegraphics[width=\linewidth]{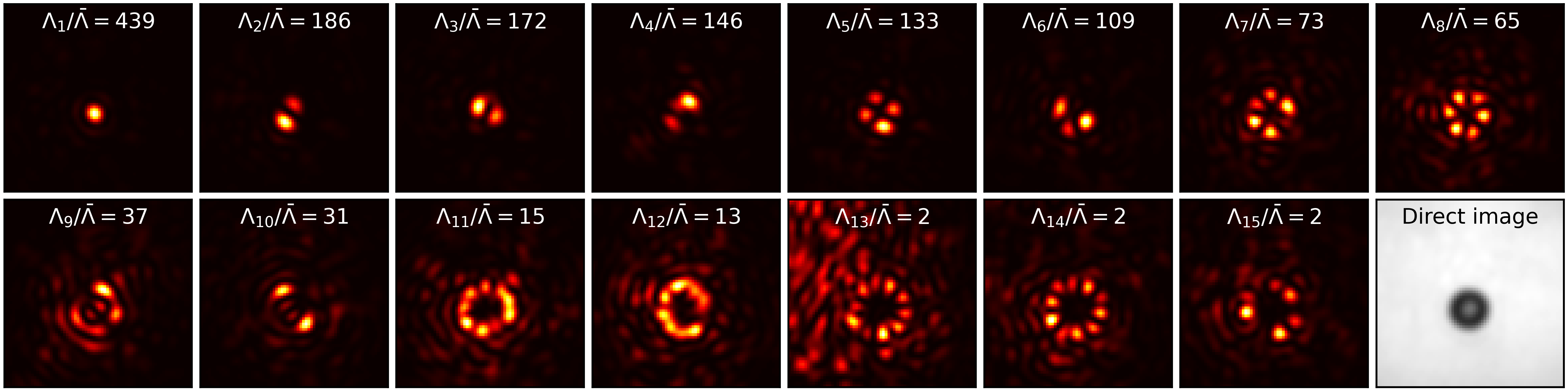}
	\end{center}
	\caption{Intensity distributions measured in the target plane for the first $15$ eigenstates of the discrimination operator $D_{12}$, along with an image of the target measured under spatially-incoherent illumination. The target is here composed of a single bead. All figures correspond to intensity distributions measured in the absence of the target, except for the last figure which is measured in the presence of the target. The field of view (edge size, 19.2\,\textmu m) is centered on the high-intensity area, and color scales are identical to those shown in Fig.~2 of the manuscript. }
	\label{fig:eigenstates_small}
\end{figure}

For a target composed of a single bead (\fig{fig:eigenstates_small}), intensity distributions have a structured aspect that resemble those of Laguerre-Gaussian modes. Assuming that light states are here optimal when they maximize the number of photons interacting with the target, these distributions can be understood as arising from a Gram-Schmidt orthogonalization procedure: the first eigenstate generate a strongly focused beam at the target position, and the $j$-th eigenstate ($j\geq2$) maximizes the number of interacting photons under the constraint that it must be orthogonal to the $j-1$ previously-calculated eigenstates. Note that the measured intensity distributions are localized around the target for the $15$ eigenstates showed in \fig{fig:eigenstates_small}, and not only for the $12$ first eigenstates whose eigenvalues are significantly above the noise level. This suggests that these last eigenstates---associated with eigenvalues $\Lambda_{13}$ to $\Lambda_{15}$---are weakly interacting with the target, at the limit of the detection capabilities of our setup.

\subsection{Extended target}

\begin{figure}[ht]
	\begin{center}
		\includegraphics[width=\linewidth]{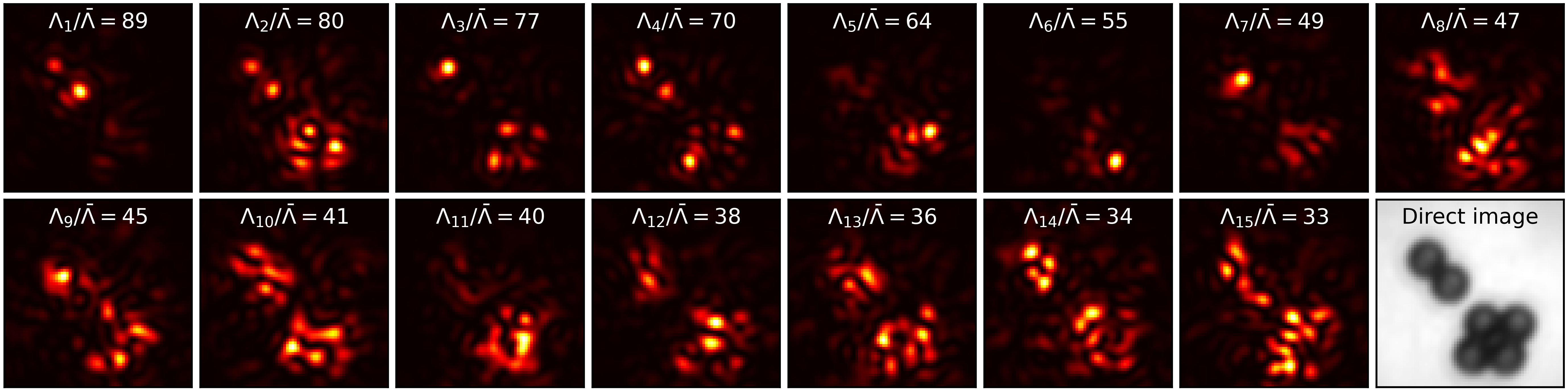}
	\end{center}
	\caption{Analogous to \fig{fig:eigenstates_small} for a target composed of $6$ beads.}
	\label{fig:eigenstates_extended}
\end{figure}

For a target composed of $6$ beads (\fig{fig:eigenstates_extended}), intensity distributions are all spatially localized around the position of the beads, but the shape of these distributions is more difficult to interpret than for the case of a single bead. For instance, the first eigenstate focuses on $2$ beads, while the third eigenstate focuses on $5$ beads and the sixth eigenstate focuses on a single bead. These different intensity distributions are likely to be due to the fact that all beads are not equally connected to the far-field modes controlled in the experiment. This clearly shows that, for complex scattering systems, the statistical distance $d_{12}$ cannot be easily maximized by a simple focusing approach. 

\begin{thebibliography}{58}%
	\makeatletter
	\providecommand \@ifxundefined [1]{%
		\@ifx{#1\undefined}
	}%
	\providecommand \@ifnum [1]{%
		\ifnum #1\expandafter \@firstoftwo
		\else \expandafter \@secondoftwo
		\fi
	}%
	\providecommand \@ifx [1]{%
		\ifx #1\expandafter \@firstoftwo
		\else \expandafter \@secondoftwo
		\fi
	}%
	\providecommand \natexlab [1]{#1}%
	\providecommand \enquote  [1]{``#1''}%
	\providecommand \bibnamefont  [1]{#1}%
	\providecommand \bibfnamefont [1]{#1}%
	\providecommand \citenamefont [1]{#1}%
	\providecommand \href@noop [0]{\@secondoftwo}%
	\providecommand \href [0]{\begingroup \@sanitize@url \@href}%
	\providecommand \@href[1]{\@@startlink{#1}\@@href}%
	\providecommand \@@href[1]{\endgroup#1\@@endlink}%
	\providecommand \@sanitize@url [0]{\catcode `\\12\catcode `\$12\catcode
		`\&12\catcode `\#12\catcode `\^12\catcode `\_12\catcode `\%12\relax}%
	\providecommand \@@startlink[1]{}%
	\providecommand \@@endlink[0]{}%
	\providecommand \url  [0]{\begingroup\@sanitize@url \@url }%
	\providecommand \@url [1]{\endgroup\@href {#1}{\urlprefix }}%
	\providecommand \urlprefix  [0]{URL }%
	\providecommand \Eprint [0]{\href }%
	\providecommand \doibase [0]{https://doi.org/}%
	\providecommand \selectlanguage [0]{\@gobble}%
	\providecommand \bibinfo  [0]{\@secondoftwo}%
	\providecommand \bibfield  [0]{\@secondoftwo}%
	\providecommand \translation [1]{[#1]}%
	\providecommand \BibitemOpen [0]{}%
	\providecommand \bibitemStop [0]{}%
	\providecommand \bibitemNoStop [0]{.\EOS\space}%
	\providecommand \EOS [0]{\spacefactor3000\relax}%
	\providecommand \BibitemShut  [1]{\csname bibitem#1\endcsname}%
	\let\auto@bib@innerbib\@empty
	%</preamble>
	\bibitem [{\citenamefont {Taylor}\ and\ \citenamefont
		{Sandoghdar}(2019)}]{taylor_interferometric_2019}%
	\BibitemOpen
	\bibfield  {author} {\bibinfo {author} {\bibfnamefont {R.~W.}\ \bibnamefont
			{Taylor}}\ and\ \bibinfo {author} {\bibfnamefont {V.}~\bibnamefont
			{Sandoghdar}},\ }\href {https://doi.org/10.1021/acs.nanolett.9b01822}
	{\bibfield  {journal} {\bibinfo  {journal} {Nano Lett.}\ }\textbf {\bibinfo
			{volume} {19}},\ \bibinfo {pages} {4827} (\bibinfo {year}
		{2019})}\BibitemShut {NoStop}%
	\bibitem [{\citenamefont {Young}\ and\ \citenamefont
		{Kukura}(2019)}]{young_interferometric_2019}%
	\BibitemOpen
	\bibfield  {author} {\bibinfo {author} {\bibfnamefont {G.}~\bibnamefont
			{Young}}\ and\ \bibinfo {author} {\bibfnamefont {P.}~\bibnamefont {Kukura}},\
	}\href {https://doi.org/10.1146/annurev-physchem-050317-021247} {\bibfield
		{journal} {\bibinfo  {journal} {Annu. Rev. Phys. Chem.}\ }\textbf {\bibinfo
			{volume} {70}},\ \bibinfo {pages} {301} (\bibinfo {year} {2019})}\BibitemShut
	{NoStop}%
	\bibitem [{\citenamefont {Orji}\ \emph {et~al.}(2018)\citenamefont {Orji},
		\citenamefont {Badaroglu}, \citenamefont {Barnes}, \citenamefont {Beitia},
		\citenamefont {Bunday}, \citenamefont {Celano}, \citenamefont {Kline},
		\citenamefont {Neisser}, \citenamefont {Obeng},\ and\ \citenamefont
		{Vladar}}]{orji_metrology_2018}%
	\BibitemOpen
	\bibfield  {author} {\bibinfo {author} {\bibfnamefont {N.~G.}\ \bibnamefont
			{Orji}}, \bibinfo {author} {\bibfnamefont {M.}~\bibnamefont {Badaroglu}},
		\bibinfo {author} {\bibfnamefont {B.~M.}\ \bibnamefont {Barnes}}, \bibinfo
		{author} {\bibfnamefont {C.}~\bibnamefont {Beitia}}, \bibinfo {author}
		{\bibfnamefont {B.~D.}\ \bibnamefont {Bunday}}, \bibinfo {author}
		{\bibfnamefont {U.}~\bibnamefont {Celano}}, \bibinfo {author} {\bibfnamefont
			{R.~J.}\ \bibnamefont {Kline}}, \bibinfo {author} {\bibfnamefont
			{M.}~\bibnamefont {Neisser}}, \bibinfo {author} {\bibfnamefont
			{Y.}~\bibnamefont {Obeng}},\ and\ \bibinfo {author} {\bibfnamefont {A.~E.}\
			\bibnamefont {Vladar}},\ }\href {https://doi.org/10.1038/s41928-018-0150-9}
	{\bibfield  {journal} {\bibinfo  {journal} {Nat. Electron.}\ }\textbf
		{\bibinfo {volume} {1}},\ \bibinfo {pages} {532} (\bibinfo {year}
		{2018})}\BibitemShut {NoStop}%
	\bibitem [{\citenamefont {Brown}\ \emph {et~al.}(2020)\citenamefont {Brown},
		\citenamefont {Brittman}, \citenamefont {Maccaferri}, \citenamefont
		{Jariwala},\ and\ \citenamefont {Celano}}]{brown_machine_2020}%
	\BibitemOpen
	\bibfield  {author} {\bibinfo {author} {\bibfnamefont {K.~A.}\ \bibnamefont
			{Brown}}, \bibinfo {author} {\bibfnamefont {S.}~\bibnamefont {Brittman}},
		\bibinfo {author} {\bibfnamefont {N.}~\bibnamefont {Maccaferri}}, \bibinfo
		{author} {\bibfnamefont {D.}~\bibnamefont {Jariwala}},\ and\ \bibinfo
		{author} {\bibfnamefont {U.}~\bibnamefont {Celano}},\ }\href
	{https://doi.org/10.1021/acs.nanolett.9b04090} {\bibfield  {journal}
		{\bibinfo  {journal} {Nano Lett.}\ }\textbf {\bibinfo {volume} {20}},\
		\bibinfo {pages} {2} (\bibinfo {year} {2020})}\BibitemShut {NoStop}%
	\bibitem [{\citenamefont {Helstrom}(1976)}]{helstrom_quantum_1976}%
	\BibitemOpen
	\bibfield  {author} {\bibinfo {author} {\bibfnamefont {C.~W.}\ \bibnamefont
			{Helstrom}},\ }\href@noop {} {\emph {\bibinfo {title} {Quantum {Detection}
				and {Estimation} {Theory}}}}\ (\bibinfo  {publisher} {Academic Press},\
	\bibinfo {address} {New York},\ \bibinfo {year} {1976})\BibitemShut {NoStop}%
	\bibitem [{\citenamefont {Weedbrook}\ \emph {et~al.}(2012)\citenamefont
		{Weedbrook}, \citenamefont {Pirandola}, \citenamefont {García-Patrón},
		\citenamefont {Cerf}, \citenamefont {Ralph}, \citenamefont {Shapiro},\ and\
		\citenamefont {Lloyd}}]{weedbrook_gaussian_2012}%
	\BibitemOpen
	\bibfield  {author} {\bibinfo {author} {\bibfnamefont {C.}~\bibnamefont
			{Weedbrook}}, \bibinfo {author} {\bibfnamefont {S.}~\bibnamefont
			{Pirandola}}, \bibinfo {author} {\bibfnamefont {R.}~\bibnamefont
			{García-Patrón}}, \bibinfo {author} {\bibfnamefont {N.~J.}\ \bibnamefont
			{Cerf}}, \bibinfo {author} {\bibfnamefont {T.~C.}\ \bibnamefont {Ralph}},
		\bibinfo {author} {\bibfnamefont {J.~H.}\ \bibnamefont {Shapiro}},\ and\
		\bibinfo {author} {\bibfnamefont {S.}~\bibnamefont {Lloyd}},\ }\href
	{https://doi.org/10.1103/RevModPhys.84.621} {\bibfield  {journal} {\bibinfo
			{journal} {Rev. Mod. Phys.}\ }\textbf {\bibinfo {volume} {84}},\ \bibinfo
		{pages} {621} (\bibinfo {year} {2012})}\BibitemShut {NoStop}%
	\bibitem [{\citenamefont {Cook}\ \emph {et~al.}(2007)\citenamefont {Cook},
		\citenamefont {Martin},\ and\ \citenamefont {Geremia}}]{cook_optical_2007}%
	\BibitemOpen
	\bibfield  {author} {\bibinfo {author} {\bibfnamefont {R.~L.}\ \bibnamefont
			{Cook}}, \bibinfo {author} {\bibfnamefont {P.~J.}\ \bibnamefont {Martin}},\
		and\ \bibinfo {author} {\bibfnamefont {J.~M.}\ \bibnamefont {Geremia}},\
	}\href {https://doi.org/10.1038/nature05655} {\bibfield  {journal} {\bibinfo
			{journal} {Nature}\ }\textbf {\bibinfo {volume} {446}},\ \bibinfo {pages}
		{774} (\bibinfo {year} {2007})}\BibitemShut {NoStop}%
	\bibitem [{\citenamefont {Becerra}\ \emph {et~al.}(2013)\citenamefont
		{Becerra}, \citenamefont {Fan}, \citenamefont {Baumgartner}, \citenamefont
		{Goldhar}, \citenamefont {Kosloski},\ and\ \citenamefont
		{Migdall}}]{becerra_experimental_2013}%
	\BibitemOpen
	\bibfield  {author} {\bibinfo {author} {\bibfnamefont {F.~E.}\ \bibnamefont
			{Becerra}}, \bibinfo {author} {\bibfnamefont {J.}~\bibnamefont {Fan}},
		\bibinfo {author} {\bibfnamefont {G.}~\bibnamefont {Baumgartner}}, \bibinfo
		{author} {\bibfnamefont {J.}~\bibnamefont {Goldhar}}, \bibinfo {author}
		{\bibfnamefont {J.~T.}\ \bibnamefont {Kosloski}},\ and\ \bibinfo {author}
		{\bibfnamefont {A.}~\bibnamefont {Migdall}},\ }\href
	{https://doi.org/10.1038/nphoton.2012.316} {\bibfield  {journal} {\bibinfo
			{journal} {Nat. Photonics}\ }\textbf {\bibinfo {volume} {7}},\ \bibinfo
		{pages} {147} (\bibinfo {year} {2013})}\BibitemShut {NoStop}%
	\bibitem [{\citenamefont {Sych}\ and\ \citenamefont
		{Leuchs}(2016)}]{sych_practical_2016}%
	\BibitemOpen
	\bibfield  {author} {\bibinfo {author} {\bibfnamefont {D.}~\bibnamefont
			{Sych}}\ and\ \bibinfo {author} {\bibfnamefont {G.}~\bibnamefont {Leuchs}},\
	}\href {https://doi.org/10.1103/PhysRevLett.117.200501} {\bibfield  {journal}
		{\bibinfo  {journal} {Phys. Rev. Lett.}\ }\textbf {\bibinfo {volume} {117}},\
		\bibinfo {pages} {200501} (\bibinfo {year} {2016})}\BibitemShut {NoStop}%
	\bibitem [{\citenamefont {Solís-Prosser}\ \emph {et~al.}(2017)\citenamefont
		{Solís-Prosser}, \citenamefont {Fernandes}, \citenamefont {Jiménez},
		\citenamefont {Delgado},\ and\ \citenamefont
		{Neves}}]{solis-prosser_experimental_2017}%
	\BibitemOpen
	\bibfield  {author} {\bibinfo {author} {\bibfnamefont {M.}~\bibnamefont
			{Solís-Prosser}}, \bibinfo {author} {\bibfnamefont {M.}~\bibnamefont
			{Fernandes}}, \bibinfo {author} {\bibfnamefont {O.}~\bibnamefont {Jiménez}},
		\bibinfo {author} {\bibfnamefont {A.}~\bibnamefont {Delgado}},\ and\ \bibinfo
		{author} {\bibfnamefont {L.}~\bibnamefont {Neves}},\ }\href
	{https://doi.org/10.1103/PhysRevLett.118.100501} {\bibfield  {journal}
		{\bibinfo  {journal} {Phys. Rev. Lett.}\ }\textbf {\bibinfo {volume} {118}},\
		\bibinfo {pages} {100501} (\bibinfo {year} {2017})}\BibitemShut {NoStop}%
	\bibitem [{\citenamefont {DiMario}\ and\ \citenamefont
		{Becerra}(2018)}]{dimario_robust_2018}%
	\BibitemOpen
	\bibfield  {author} {\bibinfo {author} {\bibfnamefont {M.}~\bibnamefont
			{DiMario}}\ and\ \bibinfo {author} {\bibfnamefont {F.}~\bibnamefont
			{Becerra}},\ }\href {https://doi.org/10.1103/PhysRevLett.121.023603}
	{\bibfield  {journal} {\bibinfo  {journal} {Phys. Rev. Lett.}\ }\textbf
		{\bibinfo {volume} {121}},\ \bibinfo {pages} {023603} (\bibinfo {year}
		{2018})}\BibitemShut {NoStop}%
	\bibitem [{\citenamefont {Lloyd}(2008)}]{lloyd_enhanced_2008}%
	\BibitemOpen
	\bibfield  {author} {\bibinfo {author} {\bibfnamefont {S.}~\bibnamefont
			{Lloyd}},\ }\href {https://doi.org/10.1126/science.1160627} {\bibfield
		{journal} {\bibinfo  {journal} {Science}\ }\textbf {\bibinfo {volume}
			{321}},\ \bibinfo {pages} {1463} (\bibinfo {year} {2008})}\BibitemShut
	{NoStop}%
	\bibitem [{\citenamefont {Pirandola}\ \emph {et~al.}(2018)\citenamefont
		{Pirandola}, \citenamefont {Bardhan}, \citenamefont {Gehring}, \citenamefont
		{Weedbrook},\ and\ \citenamefont {Lloyd}}]{pirandola_advances_2018}%
	\BibitemOpen
	\bibfield  {author} {\bibinfo {author} {\bibfnamefont {S.}~\bibnamefont
			{Pirandola}}, \bibinfo {author} {\bibfnamefont {B.~R.}\ \bibnamefont
			{Bardhan}}, \bibinfo {author} {\bibfnamefont {T.}~\bibnamefont {Gehring}},
		\bibinfo {author} {\bibfnamefont {C.}~\bibnamefont {Weedbrook}},\ and\
		\bibinfo {author} {\bibfnamefont {S.}~\bibnamefont {Lloyd}},\ }\href
	{https://doi.org/10.1038/s41566-018-0301-6} {\bibfield  {journal} {\bibinfo
			{journal} {Nat. Photonics}\ }\textbf {\bibinfo {volume} {12}},\ \bibinfo
		{pages} {724} (\bibinfo {year} {2018})}\BibitemShut {NoStop}%
	\bibitem [{\citenamefont {Nair}\ and\ \citenamefont
		{Gu}(2020)}]{nair_fundamental_2020}%
	\BibitemOpen
	\bibfield  {author} {\bibinfo {author} {\bibfnamefont {R.}~\bibnamefont
			{Nair}}\ and\ \bibinfo {author} {\bibfnamefont {M.}~\bibnamefont {Gu}},\
	}\href {https://doi.org/10.1364/OPTICA.391335} {\bibfield  {journal}
		{\bibinfo  {journal} {Optica}\ }\textbf {\bibinfo {volume} {7}},\ \bibinfo
		{pages} {771} (\bibinfo {year} {2020})}\BibitemShut {NoStop}%
	\bibitem [{\citenamefont {Mosk}\ \emph {et~al.}(2012)\citenamefont {Mosk},
		\citenamefont {Lagendijk}, \citenamefont {Lerosey},\ and\ \citenamefont
		{Fink}}]{mosk_controlling_2012}%
	\BibitemOpen
	\bibfield  {author} {\bibinfo {author} {\bibfnamefont {A.~P.}\ \bibnamefont
			{Mosk}}, \bibinfo {author} {\bibfnamefont {A.}~\bibnamefont {Lagendijk}},
		\bibinfo {author} {\bibfnamefont {G.}~\bibnamefont {Lerosey}},\ and\ \bibinfo
		{author} {\bibfnamefont {M.}~\bibnamefont {Fink}},\ }\href
	{https://doi.org/10.1038/nphoton.2012.88} {\bibfield  {journal} {\bibinfo
			{journal} {Nat. Photonics}\ }\textbf {\bibinfo {volume} {6}},\ \bibinfo
		{pages} {283} (\bibinfo {year} {2012})}\BibitemShut {NoStop}%
	\bibitem [{\citenamefont {Horstmeyer}\ \emph {et~al.}(2015)\citenamefont
		{Horstmeyer}, \citenamefont {Ruan},\ and\ \citenamefont
		{Yang}}]{horstmeyer_guidestar-assisted_2015}%
	\BibitemOpen
	\bibfield  {author} {\bibinfo {author} {\bibfnamefont {R.}~\bibnamefont
			{Horstmeyer}}, \bibinfo {author} {\bibfnamefont {H.}~\bibnamefont {Ruan}},\
		and\ \bibinfo {author} {\bibfnamefont {C.}~\bibnamefont {Yang}},\ }\href
	{https://doi.org/10.1038/nphoton.2015.140} {\bibfield  {journal} {\bibinfo
			{journal} {Nat. Photonics}\ }\textbf {\bibinfo {volume} {9}},\ \bibinfo
		{pages} {563} (\bibinfo {year} {2015})}\BibitemShut {NoStop}%
	\bibitem [{\citenamefont {Rotter}\ and\ \citenamefont
		{Gigan}(2017)}]{rotter_light_2017}%
	\BibitemOpen
	\bibfield  {author} {\bibinfo {author} {\bibfnamefont {S.}~\bibnamefont
			{Rotter}}\ and\ \bibinfo {author} {\bibfnamefont {S.}~\bibnamefont {Gigan}},\
	}\href {https://doi.org/10.1103/RevModPhys.89.015005} {\bibfield  {journal}
		{\bibinfo  {journal} {Rev. Mod. Phys.}\ }\textbf {\bibinfo {volume} {89}},\
		\bibinfo {pages} {015005} (\bibinfo {year} {2017})}\BibitemShut {NoStop}%
	\bibitem [{\citenamefont {Vellekoop}\ and\ \citenamefont
		{Mosk}(2008)}]{vellekoop_universal_2008}%
	\BibitemOpen
	\bibfield  {author} {\bibinfo {author} {\bibfnamefont {I.~M.}\ \bibnamefont
			{Vellekoop}}\ and\ \bibinfo {author} {\bibfnamefont {A.~P.}\ \bibnamefont
			{Mosk}},\ }\href {https://doi.org/10.1103/PhysRevLett.101.120601} {\bibfield
		{journal} {\bibinfo  {journal} {Phys. Rev. Lett.}\ }\textbf {\bibinfo
			{volume} {101}},\ \bibinfo {pages} {120601} (\bibinfo {year}
		{2008})}\BibitemShut {NoStop}%
	\bibitem [{\citenamefont {Kim}\ \emph {et~al.}(2012)\citenamefont {Kim},
		\citenamefont {Choi}, \citenamefont {Yoon}, \citenamefont {Choi},
		\citenamefont {Kim}, \citenamefont {Park},\ and\ \citenamefont
		{Choi}}]{kim_maximal_2012}%
	\BibitemOpen
	\bibfield  {author} {\bibinfo {author} {\bibfnamefont {M.}~\bibnamefont
			{Kim}}, \bibinfo {author} {\bibfnamefont {Y.}~\bibnamefont {Choi}}, \bibinfo
		{author} {\bibfnamefont {C.}~\bibnamefont {Yoon}}, \bibinfo {author}
		{\bibfnamefont {W.}~\bibnamefont {Choi}}, \bibinfo {author} {\bibfnamefont
			{J.}~\bibnamefont {Kim}}, \bibinfo {author} {\bibfnamefont {Q.-H.}\
			\bibnamefont {Park}},\ and\ \bibinfo {author} {\bibfnamefont
			{W.}~\bibnamefont {Choi}},\ }\href {https://doi.org/10.1038/nphoton.2012.159}
	{\bibfield  {journal} {\bibinfo  {journal} {Nat. Photonics}\ }\textbf
		{\bibinfo {volume} {6}},\ \bibinfo {pages} {581} (\bibinfo {year}
		{2012})}\BibitemShut {NoStop}%
	\bibitem [{\citenamefont {Popoff}\ \emph {et~al.}(2014)\citenamefont {Popoff},
		\citenamefont {Goetschy}, \citenamefont {Liew}, \citenamefont {Stone},\ and\
		\citenamefont {Cao}}]{popoff_coherent_2014}%
	\BibitemOpen
	\bibfield  {author} {\bibinfo {author} {\bibfnamefont {S.}~\bibnamefont
			{Popoff}}, \bibinfo {author} {\bibfnamefont {A.}~\bibnamefont {Goetschy}},
		\bibinfo {author} {\bibfnamefont {S.}~\bibnamefont {Liew}}, \bibinfo {author}
		{\bibfnamefont {A.}~\bibnamefont {Stone}},\ and\ \bibinfo {author}
		{\bibfnamefont {H.}~\bibnamefont {Cao}},\ }\href
	{https://doi.org/10.1103/PhysRevLett.112.133903} {\bibfield  {journal}
		{\bibinfo  {journal} {Phys. Rev. Lett.}\ }\textbf {\bibinfo {volume} {112}},\
		\bibinfo {pages} {133903} (\bibinfo {year} {2014})}\BibitemShut {NoStop}%
	\bibitem [{\citenamefont {Cheng}\ and\ \citenamefont
		{Genack}(2014)}]{cheng_focusing_2014}%
	\BibitemOpen
	\bibfield  {author} {\bibinfo {author} {\bibfnamefont {X.}~\bibnamefont
			{Cheng}}\ and\ \bibinfo {author} {\bibfnamefont {A.~Z.}\ \bibnamefont
			{Genack}},\ }\href {https://doi.org/10.1364/OL.39.006324} {\bibfield
		{journal} {\bibinfo  {journal} {Opt. Lett.}\ }\textbf {\bibinfo {volume}
			{39}},\ \bibinfo {pages} {6324} (\bibinfo {year} {2014})}\BibitemShut
	{NoStop}%
	\bibitem [{\citenamefont {Ambichl}\ \emph {et~al.}(2017)\citenamefont
		{Ambichl}, \citenamefont {Brandstötter}, \citenamefont {Böhm},
		\citenamefont {Kühmayer}, \citenamefont {Kuhl},\ and\ \citenamefont
		{Rotter}}]{ambichl_focusing_2017}%
	\BibitemOpen
	\bibfield  {author} {\bibinfo {author} {\bibfnamefont {P.}~\bibnamefont
			{Ambichl}}, \bibinfo {author} {\bibfnamefont {A.}~\bibnamefont
			{Brandstötter}}, \bibinfo {author} {\bibfnamefont {J.}~\bibnamefont
			{Böhm}}, \bibinfo {author} {\bibfnamefont {M.}~\bibnamefont {Kühmayer}},
		\bibinfo {author} {\bibfnamefont {U.}~\bibnamefont {Kuhl}},\ and\ \bibinfo
		{author} {\bibfnamefont {S.}~\bibnamefont {Rotter}},\ }\href
	{https://doi.org/10.1103/PhysRevLett.119.033903} {\bibfield  {journal}
		{\bibinfo  {journal} {Phys. Rev. Lett.}\ }\textbf {\bibinfo {volume} {119}},\
		\bibinfo {pages} {033903} (\bibinfo {year} {2017})}\BibitemShut {NoStop}%
	\bibitem [{\citenamefont {Durand}\ \emph {et~al.}(2019)\citenamefont {Durand},
		\citenamefont {Popoff}, \citenamefont {Carminati},\ and\ \citenamefont
		{Goetschy}}]{durand_optimizing_2019}%
	\BibitemOpen
	\bibfield  {author} {\bibinfo {author} {\bibfnamefont {M.}~\bibnamefont
			{Durand}}, \bibinfo {author} {\bibfnamefont {S.}~\bibnamefont {Popoff}},
		\bibinfo {author} {\bibfnamefont {R.}~\bibnamefont {Carminati}},\ and\
		\bibinfo {author} {\bibfnamefont {A.}~\bibnamefont {Goetschy}},\ }\href
	{https://doi.org/10.1103/PhysRevLett.123.243901} {\bibfield  {journal}
		{\bibinfo  {journal} {Phys. Rev. Lett.}\ }\textbf {\bibinfo {volume} {123}},\
		\bibinfo {pages} {243901} (\bibinfo {year} {2019})}\BibitemShut {NoStop}%
	\bibitem [{\citenamefont {Bender}\ \emph {et~al.}(2021)\citenamefont {Bender},
		\citenamefont {Yamilov}, \citenamefont {Goetschy}, \citenamefont {Yilmaz},
		\citenamefont {Hsu},\ and\ \citenamefont {Cao}}]{bender_depth-targeted_2021}%
	\BibitemOpen
	\bibfield  {author} {\bibinfo {author} {\bibfnamefont {N.}~\bibnamefont
			{Bender}}, \bibinfo {author} {\bibfnamefont {A.}~\bibnamefont {Yamilov}},
		\bibinfo {author} {\bibfnamefont {A.}~\bibnamefont {Goetschy}}, \bibinfo
		{author} {\bibfnamefont {H.}~\bibnamefont {Yilmaz}}, \bibinfo {author}
		{\bibfnamefont {C.~W.}\ \bibnamefont {Hsu}},\ and\ \bibinfo {author}
		{\bibfnamefont {H.}~\bibnamefont {Cao}},\ }\href
	{http://arxiv.org/abs/2105.13417} {\bibfield  {journal} {\bibinfo  {journal}
			{arXiv:2105.13417}\ } (\bibinfo {year} {2021})}\BibitemShut {NoStop}%
	\bibitem [{\citenamefont {Bouchet}\ \emph {et~al.}(2020)\citenamefont
		{Bouchet}, \citenamefont {Carminati},\ and\ \citenamefont
		{Mosk}}]{bouchet_influence_2020}%
	\BibitemOpen
	\bibfield  {author} {\bibinfo {author} {\bibfnamefont {D.}~\bibnamefont
			{Bouchet}}, \bibinfo {author} {\bibfnamefont {R.}~\bibnamefont {Carminati}},\
		and\ \bibinfo {author} {\bibfnamefont {A.~P.}\ \bibnamefont {Mosk}},\ }\href
	{https://doi.org/10.1103/PhysRevLett.124.133903} {\bibfield  {journal}
		{\bibinfo  {journal} {Phys. Rev. Lett.}\ }\textbf {\bibinfo {volume} {124}},\
		\bibinfo {pages} {133903} (\bibinfo {year} {2020})}\BibitemShut {NoStop}%
	\bibitem [{\citenamefont {Bouchet}\ \emph
		{et~al.}(2021{\natexlab{a}})\citenamefont {Bouchet}, \citenamefont {Rotter},\
		and\ \citenamefont {Mosk}}]{bouchet_maximum_2021}%
	\BibitemOpen
	\bibfield  {author} {\bibinfo {author} {\bibfnamefont {D.}~\bibnamefont
			{Bouchet}}, \bibinfo {author} {\bibfnamefont {S.}~\bibnamefont {Rotter}},\
		and\ \bibinfo {author} {\bibfnamefont {A.~P.}\ \bibnamefont {Mosk}},\ }\href
	{https://doi.org/10.1038/s41567-020-01137-4} {\bibfield  {journal} {\bibinfo
			{journal} {Nat. Phys.}\ }\textbf {\bibinfo {volume} {17}},\ \bibinfo {pages}
		{564} (\bibinfo {year} {2021}{\natexlab{a}})}\BibitemShut {NoStop}%
	\bibitem [{\citenamefont {Hershko}\ \emph {et~al.}(2019)\citenamefont
		{Hershko}, \citenamefont {Weiss}, \citenamefont {Michaeli},\ and\
		\citenamefont {Shechtman}}]{hershko_multicolor_2019}%
	\BibitemOpen
	\bibfield  {author} {\bibinfo {author} {\bibfnamefont {E.}~\bibnamefont
			{Hershko}}, \bibinfo {author} {\bibfnamefont {L.~E.}\ \bibnamefont {Weiss}},
		\bibinfo {author} {\bibfnamefont {T.}~\bibnamefont {Michaeli}},\ and\
		\bibinfo {author} {\bibfnamefont {Y.}~\bibnamefont {Shechtman}},\ }\href
	{https://doi.org/10.1364/OE.27.006158} {\bibfield  {journal} {\bibinfo
			{journal} {Opt. Express}\ }\textbf {\bibinfo {volume} {27}},\ \bibinfo
		{pages} {6158} (\bibinfo {year} {2019})}\BibitemShut {NoStop}%
	\bibitem [{\citenamefont {Kellman}\ \emph {et~al.}(2019)\citenamefont
		{Kellman}, \citenamefont {Bostan}, \citenamefont {Repina},\ and\
		\citenamefont {Waller}}]{kellman_physics-based_2019}%
	\BibitemOpen
	\bibfield  {author} {\bibinfo {author} {\bibfnamefont {M.~R.}\ \bibnamefont
			{Kellman}}, \bibinfo {author} {\bibfnamefont {E.}~\bibnamefont {Bostan}},
		\bibinfo {author} {\bibfnamefont {N.~A.}\ \bibnamefont {Repina}},\ and\
		\bibinfo {author} {\bibfnamefont {L.}~\bibnamefont {Waller}},\ }\href
	{https://doi.org/10.1109/TCI.2019.2905434} {\bibfield  {journal} {\bibinfo
			{journal} {IEEE Trans. Comput. Imaging}\ }\textbf {\bibinfo {volume} {5}},\
		\bibinfo {pages} {344} (\bibinfo {year} {2019})}\BibitemShut {NoStop}%
	\bibitem [{\citenamefont {Muthumbi}\ \emph {et~al.}(2019)\citenamefont
		{Muthumbi}, \citenamefont {Chaware}, \citenamefont {Kim}, \citenamefont
		{Zhou}, \citenamefont {Konda}, \citenamefont {Chen}, \citenamefont
		{Judkewitz}, \citenamefont {Erdmann}, \citenamefont {Kappes},\ and\
		\citenamefont {Horstmeyer}}]{muthumbi_learned_2019}%
	\BibitemOpen
	\bibfield  {author} {\bibinfo {author} {\bibfnamefont {A.}~\bibnamefont
			{Muthumbi}}, \bibinfo {author} {\bibfnamefont {A.}~\bibnamefont {Chaware}},
		\bibinfo {author} {\bibfnamefont {K.}~\bibnamefont {Kim}}, \bibinfo {author}
		{\bibfnamefont {K.~C.}\ \bibnamefont {Zhou}}, \bibinfo {author}
		{\bibfnamefont {P.~C.}\ \bibnamefont {Konda}}, \bibinfo {author}
		{\bibfnamefont {R.}~\bibnamefont {Chen}}, \bibinfo {author} {\bibfnamefont
			{B.}~\bibnamefont {Judkewitz}}, \bibinfo {author} {\bibfnamefont
			{A.}~\bibnamefont {Erdmann}}, \bibinfo {author} {\bibfnamefont
			{B.}~\bibnamefont {Kappes}},\ and\ \bibinfo {author} {\bibfnamefont
			{R.}~\bibnamefont {Horstmeyer}},\ }\href
	{https://doi.org/10.1364/BOE.10.006351} {\bibfield  {journal} {\bibinfo
			{journal} {Biomed. Opt. Express}\ }\textbf {\bibinfo {volume} {10}},\
		\bibinfo {pages} {6351} (\bibinfo {year} {2019})}\BibitemShut {NoStop}%
	\bibitem [{\citenamefont {Hougne}\ \emph {et~al.}(2020)\citenamefont {Hougne},
		\citenamefont {Imani}, \citenamefont {Diebold}, \citenamefont {Horstmeyer},\
		and\ \citenamefont {Smith}}]{hougne_learned_2020}%
	\BibitemOpen
	\bibfield  {author} {\bibinfo {author} {\bibfnamefont {P.~d.}\ \bibnamefont
			{Hougne}}, \bibinfo {author} {\bibfnamefont {M.~F.}\ \bibnamefont {Imani}},
		\bibinfo {author} {\bibfnamefont {A.~V.}\ \bibnamefont {Diebold}}, \bibinfo
		{author} {\bibfnamefont {R.}~\bibnamefont {Horstmeyer}},\ and\ \bibinfo
		{author} {\bibfnamefont {D.~R.}\ \bibnamefont {Smith}},\ }\href
	{https://doi.org/10.1002/advs.201901913} {\bibfield  {journal} {\bibinfo
			{journal} {Adv. Sci.}\ }\textbf {\bibinfo {volume} {7}},\ \bibinfo {pages}
		{1901913} (\bibinfo {year} {2020})}\BibitemShut {NoStop}%
	\bibitem [{\citenamefont {Horisaki}\ \emph {et~al.}(2020)\citenamefont
		{Horisaki}, \citenamefont {Horisaki}, \citenamefont {Okamoto},\ and\
		\citenamefont {Tanida}}]{horisaki_deeply_2020}%
	\BibitemOpen
	\bibfield  {author} {\bibinfo {author} {\bibfnamefont {R.}~\bibnamefont
			{Horisaki}}, \bibinfo {author} {\bibfnamefont {R.}~\bibnamefont {Horisaki}},
		\bibinfo {author} {\bibfnamefont {Y.}~\bibnamefont {Okamoto}},\ and\ \bibinfo
		{author} {\bibfnamefont {J.}~\bibnamefont {Tanida}},\ }\href
	{https://doi.org/10.1364/OL.390810} {\bibfield  {journal} {\bibinfo
			{journal} {Opt. Lett.}\ }\textbf {\bibinfo {volume} {45}},\ \bibinfo {pages}
		{3131} (\bibinfo {year} {2020})}\BibitemShut {NoStop}%
	\bibitem [{sup()}]{supplementary}%
	\BibitemOpen
	\href@noop {} {}\bibinfo {note} {See Supplemental Material for
		detailed derivations and additional experimental results. Supplemental
		Material includes
		Refs.~\cite{loudon_quantum_2000,goodman_statistical_2015,pai_optical_2020,mirhosseini_rapid_2013}}\BibitemShut
	{NoStop}%
	\bibitem [{\citenamefont {Popoff}\ \emph {et~al.}(2010)\citenamefont {Popoff},
		\citenamefont {Lerosey}, \citenamefont {Carminati}, \citenamefont {Fink},
		\citenamefont {Boccara},\ and\ \citenamefont
		{Gigan}}]{popoff_measuring_2010}%
	\BibitemOpen
	\bibfield  {author} {\bibinfo {author} {\bibfnamefont {S.~M.}\ \bibnamefont
			{Popoff}}, \bibinfo {author} {\bibfnamefont {G.}~\bibnamefont {Lerosey}},
		\bibinfo {author} {\bibfnamefont {R.}~\bibnamefont {Carminati}}, \bibinfo
		{author} {\bibfnamefont {M.}~\bibnamefont {Fink}}, \bibinfo {author}
		{\bibfnamefont {A.~C.}\ \bibnamefont {Boccara}},\ and\ \bibinfo {author}
		{\bibfnamefont {S.}~\bibnamefont {Gigan}},\ }\href
	{https://doi.org/10.1103/PhysRevLett.104.100601} {\bibfield  {journal}
		{\bibinfo  {journal} {Phys. Rev. Lett.}\ }\textbf {\bibinfo {volume} {104}},\
		\bibinfo {pages} {100601} (\bibinfo {year} {2010})}\BibitemShut {NoStop}%
	\bibitem [{\citenamefont {Lee}(1974)}]{lee_binary_1974}%
	\BibitemOpen
	\bibfield  {author} {\bibinfo {author} {\bibfnamefont {W.-H.}\ \bibnamefont
			{Lee}},\ }\href {https://doi.org/10.1364/AO.13.001677} {\bibfield  {journal}
		{\bibinfo  {journal} {Appl. Opt.}\ }\textbf {\bibinfo {volume} {13}},\
		\bibinfo {pages} {1677} (\bibinfo {year} {1974})}\BibitemShut {NoStop}%
	\bibitem [{\citenamefont {Cuche}\ \emph {et~al.}(2000)\citenamefont {Cuche},
		\citenamefont {Marquet},\ and\ \citenamefont
		{Depeursinge}}]{cuche_spatial_2000}%
	\BibitemOpen
	\bibfield  {author} {\bibinfo {author} {\bibfnamefont {E.}~\bibnamefont
			{Cuche}}, \bibinfo {author} {\bibfnamefont {P.}~\bibnamefont {Marquet}},\
		and\ \bibinfo {author} {\bibfnamefont {C.}~\bibnamefont {Depeursinge}},\
	}\href {https://doi.org/10.1364/AO.39.004070} {\bibfield  {journal} {\bibinfo
			{journal} {Appl. Opt.}\ }\textbf {\bibinfo {volume} {39}},\ \bibinfo {pages}
		{4070} (\bibinfo {year} {2000})}\BibitemShut {NoStop}%
	\bibitem [{\citenamefont {Takeoka}\ and\ \citenamefont
		{Sasaki}(2008)}]{takeoka_discrimination_2008}%
	\BibitemOpen
	\bibfield  {author} {\bibinfo {author} {\bibfnamefont {M.}~\bibnamefont
			{Takeoka}}\ and\ \bibinfo {author} {\bibfnamefont {M.}~\bibnamefont
			{Sasaki}},\ }\href {https://doi.org/10.1103/PhysRevA.78.022320} {\bibfield
		{journal} {\bibinfo  {journal} {Phys. Rev. A}\ }\textbf {\bibinfo {volume}
			{78}},\ \bibinfo {pages} {022320} (\bibinfo {year} {2008})}\BibitemShut
	{NoStop}%
	\bibitem [{\citenamefont {Trees}\ \emph {et~al.}(2013)\citenamefont {Trees},
		\citenamefont {Bell},\ and\ \citenamefont {Tian}}]{trees_detection_2013}%
	\BibitemOpen
	\bibfield  {author} {\bibinfo {author} {\bibfnamefont {H.~L.~V.}\
			\bibnamefont {Trees}}, \bibinfo {author} {\bibfnamefont {K.~L.}\ \bibnamefont
			{Bell}},\ and\ \bibinfo {author} {\bibfnamefont {Z.}~\bibnamefont {Tian}},\
	}\href@noop {} {\emph {\bibinfo {title} {Detection {Estimation} and
				{Modulation} {Theory}, {Part} {I}}}}\ (\bibinfo  {publisher} {John Wiley \&
		Sons},\ \bibinfo {address} {Hoboken},\ \bibinfo {year} {2013})\BibitemShut
	{NoStop}%
	\bibitem [{\citenamefont {Yu}\ \emph {et~al.}(2013)\citenamefont {Yu},
		\citenamefont {Hillman}, \citenamefont {Choi}, \citenamefont {Lee},
		\citenamefont {Feld}, \citenamefont {Dasari},\ and\ \citenamefont
		{Park}}]{yu_measuring_2013}%
	\BibitemOpen
	\bibfield  {author} {\bibinfo {author} {\bibfnamefont {H.}~\bibnamefont
			{Yu}}, \bibinfo {author} {\bibfnamefont {T.~R.}\ \bibnamefont {Hillman}},
		\bibinfo {author} {\bibfnamefont {W.}~\bibnamefont {Choi}}, \bibinfo {author}
		{\bibfnamefont {J.~O.}\ \bibnamefont {Lee}}, \bibinfo {author} {\bibfnamefont
			{M.~S.}\ \bibnamefont {Feld}}, \bibinfo {author} {\bibfnamefont {R.~R.}\
			\bibnamefont {Dasari}},\ and\ \bibinfo {author} {\bibfnamefont
			{Y.}~\bibnamefont {Park}},\ }\href
	{https://doi.org/10.1103/PhysRevLett.111.153902} {\bibfield  {journal}
		{\bibinfo  {journal} {Phys. Rev. Lett.}\ }\textbf {\bibinfo {volume} {111}},\
		\bibinfo {pages} {153902} (\bibinfo {year} {2013})}\BibitemShut {NoStop}%
	\bibitem [{\citenamefont {Plöschner}\ \emph {et~al.}(2015)\citenamefont
		{Plöschner}, \citenamefont {Tyc},\ and\ \citenamefont
		{Čižmár}}]{ploschner_seeing_2015}%
	\BibitemOpen
	\bibfield  {author} {\bibinfo {author} {\bibfnamefont {M.}~\bibnamefont
			{Plöschner}}, \bibinfo {author} {\bibfnamefont {T.}~\bibnamefont {Tyc}},\
		and\ \bibinfo {author} {\bibfnamefont {T.}~\bibnamefont {Čižmár}},\ }\href
	{https://doi.org/10.1038/nphoton.2015.112} {\bibfield  {journal} {\bibinfo
			{journal} {Nat. Photonics}\ }\textbf {\bibinfo {volume} {9}},\ \bibinfo
		{pages} {529} (\bibinfo {year} {2015})}\BibitemShut {NoStop}%
	\bibitem [{\citenamefont {Matthès}\ \emph {et~al.}(2021)\citenamefont
		{Matthès}, \citenamefont {Bromberg}, \citenamefont {de~Rosny},\ and\
		\citenamefont {Popoff}}]{matthes_learning_2021}%
	\BibitemOpen
	\bibfield  {author} {\bibinfo {author} {\bibfnamefont {M.~W.}\ \bibnamefont
			{Matthès}}, \bibinfo {author} {\bibfnamefont {Y.}~\bibnamefont {Bromberg}},
		\bibinfo {author} {\bibfnamefont {J.}~\bibnamefont {de~Rosny}},\ and\
		\bibinfo {author} {\bibfnamefont {S.~M.}\ \bibnamefont {Popoff}},\ }\href
	{https://doi.org/10.1103/PhysRevX.11.021060} {\bibfield  {journal} {\bibinfo
			{journal} {Phys. Rev. X}\ }\textbf {\bibinfo {volume} {11}},\ \bibinfo
		{pages} {021060} (\bibinfo {year} {2021})}\BibitemShut {NoStop}%
	\bibitem [{\citenamefont {Pai}\ \emph {et~al.}(2021)\citenamefont {Pai},
		\citenamefont {Bosch}, \citenamefont {Kühmayer}, \citenamefont {Rotter},\
		and\ \citenamefont {Mosk}}]{pai_scattering_2021}%
	\BibitemOpen
	\bibfield  {author} {\bibinfo {author} {\bibfnamefont {P.}~\bibnamefont
			{Pai}}, \bibinfo {author} {\bibfnamefont {J.}~\bibnamefont {Bosch}}, \bibinfo
		{author} {\bibfnamefont {M.}~\bibnamefont {Kühmayer}}, \bibinfo {author}
		{\bibfnamefont {S.}~\bibnamefont {Rotter}},\ and\ \bibinfo {author}
		{\bibfnamefont {A.~P.}\ \bibnamefont {Mosk}},\ }\href
	{https://doi.org/10.1038/s41566-021-00789-9} {\bibfield  {journal} {\bibinfo
			{journal} {Nat. Photonics}\ }\textbf {\bibinfo {volume} {15}},\ \bibinfo
		{pages} {431} (\bibinfo {year} {2021})}\BibitemShut {NoStop}%
	\bibitem [{\citenamefont {Bag}\ \emph {et~al.}(2020)\citenamefont {Bag},
		\citenamefont {Neugebauer}, \citenamefont {Mick}, \citenamefont
		{Christiansen}, \citenamefont {Schulz},\ and\ \citenamefont
		{Banzer}}]{bag_towards_2020}%
	\BibitemOpen
	\bibfield  {author} {\bibinfo {author} {\bibfnamefont {A.}~\bibnamefont
			{Bag}}, \bibinfo {author} {\bibfnamefont {M.}~\bibnamefont {Neugebauer}},
		\bibinfo {author} {\bibfnamefont {U.}~\bibnamefont {Mick}}, \bibinfo {author}
		{\bibfnamefont {S.}~\bibnamefont {Christiansen}}, \bibinfo {author}
		{\bibfnamefont {S.~A.}\ \bibnamefont {Schulz}},\ and\ \bibinfo {author}
		{\bibfnamefont {P.}~\bibnamefont {Banzer}},\ }\href
	{https://doi.org/10.1038/s41467-020-16739-y} {\bibfield  {journal} {\bibinfo
			{journal} {Nat. Commun.}\ }\textbf {\bibinfo {volume} {11}},\ \bibinfo
		{pages} {2915} (\bibinfo {year} {2020})}\BibitemShut {NoStop}%
	\bibitem [{\citenamefont {Wolterink}\ \emph {et~al.}(2021)\citenamefont
		{Wolterink}, \citenamefont {Buijs}, \citenamefont {Gerini}, \citenamefont
		{Koenderink},\ and\ \citenamefont {Verhagen}}]{wolterink_localizing_2021}%
	\BibitemOpen
	\bibfield  {author} {\bibinfo {author} {\bibfnamefont {T.~A.~W.}\
			\bibnamefont {Wolterink}}, \bibinfo {author} {\bibfnamefont {R.~D.}\
			\bibnamefont {Buijs}}, \bibinfo {author} {\bibfnamefont {G.}~\bibnamefont
			{Gerini}}, \bibinfo {author} {\bibfnamefont {A.~F.}\ \bibnamefont
			{Koenderink}},\ and\ \bibinfo {author} {\bibfnamefont {E.}~\bibnamefont
			{Verhagen}},\ }\href {https://doi.org/10.1515/nanoph-2020-0669} {\bibfield
		{journal} {\bibinfo  {journal} {Nanophotonics}\ }\textbf {\bibinfo {volume}
			{10}},\ \bibinfo {pages} {1723} (\bibinfo {year} {2021})}\BibitemShut
	{NoStop}%
	\bibitem [{\citenamefont {Buijs}\ \emph {et~al.}(2021)\citenamefont {Buijs},
		\citenamefont {Wolterink}, \citenamefont {Gerini}, \citenamefont {Verhagen},\
		and\ \citenamefont {Koenderink}}]{buijs_programming_2021}%
	\BibitemOpen
	\bibfield  {author} {\bibinfo {author} {\bibfnamefont {R.~D.}\ \bibnamefont
			{Buijs}}, \bibinfo {author} {\bibfnamefont {T.~A.~W.}\ \bibnamefont
			{Wolterink}}, \bibinfo {author} {\bibfnamefont {G.}~\bibnamefont {Gerini}},
		\bibinfo {author} {\bibfnamefont {E.}~\bibnamefont {Verhagen}},\ and\
		\bibinfo {author} {\bibfnamefont {A.~F.}\ \bibnamefont {Koenderink}},\ }\href
	{https://doi.org/10.1002/adom.202100435} {\bibfield  {journal} {\bibinfo
			{journal} {Adv. Opt. Mater.}\ }\textbf {\bibinfo {volume} {9}},\ \bibinfo
		{pages} {2100435} (\bibinfo {year} {2021})}\BibitemShut {NoStop}%
	\bibitem [{\citenamefont {del Hougne}\ \emph {et~al.}(2021)\citenamefont {del
			Hougne}, \citenamefont {Gigan},\ and\ \citenamefont {del
			Hougne}}]{del_hougne_deeply_2021}%
	\BibitemOpen
	\bibfield  {author} {\bibinfo {author} {\bibfnamefont {M.}~\bibnamefont {del
				Hougne}}, \bibinfo {author} {\bibfnamefont {S.}~\bibnamefont {Gigan}},\ and\
		\bibinfo {author} {\bibfnamefont {P.}~\bibnamefont {del Hougne}},\ }\href
	{https://doi.org/10.1103/PhysRevLett.127.043903} {\bibfield  {journal}
		{\bibinfo  {journal} {Phys. Rev. Lett.}\ }\textbf {\bibinfo {volume} {127}},\
		\bibinfo {pages} {043903} (\bibinfo {year} {2021})}\BibitemShut {NoStop}%
	\bibitem [{\citenamefont {Fink}\ \emph {et~al.}(2000)\citenamefont {Fink},
		\citenamefont {Cassereau}, \citenamefont {Derode}, \citenamefont {Prada},
		\citenamefont {Roux}, \citenamefont {Tanter}, \citenamefont {Thomas},\ and\
		\citenamefont {Wu}}]{fink_time-reversed_2000}%
	\BibitemOpen
	\bibfield  {author} {\bibinfo {author} {\bibfnamefont {M.}~\bibnamefont
			{Fink}}, \bibinfo {author} {\bibfnamefont {D.}~\bibnamefont {Cassereau}},
		\bibinfo {author} {\bibfnamefont {A.}~\bibnamefont {Derode}}, \bibinfo
		{author} {\bibfnamefont {C.}~\bibnamefont {Prada}}, \bibinfo {author}
		{\bibfnamefont {P.}~\bibnamefont {Roux}}, \bibinfo {author} {\bibfnamefont
			{M.}~\bibnamefont {Tanter}}, \bibinfo {author} {\bibfnamefont {J.-L.}\
			\bibnamefont {Thomas}},\ and\ \bibinfo {author} {\bibfnamefont
			{F.}~\bibnamefont {Wu}},\ }\href
	{https://doi.org/10.1088/0034-4885/63/12/202} {\bibfield  {journal} {\bibinfo
			{journal} {Rep. Prog. Phys.}\ }\textbf {\bibinfo {volume} {63}},\ \bibinfo
		{pages} {1933} (\bibinfo {year} {2000})}\BibitemShut {NoStop}%
	\bibitem [{\citenamefont {Zhou}\ \emph {et~al.}(2014)\citenamefont {Zhou},
		\citenamefont {Ruan}, \citenamefont {Yang},\ and\ \citenamefont
		{Judkewitz}}]{zhou_focusing_2014}%
	\BibitemOpen
	\bibfield  {author} {\bibinfo {author} {\bibfnamefont {E.~H.}\ \bibnamefont
			{Zhou}}, \bibinfo {author} {\bibfnamefont {H.}~\bibnamefont {Ruan}}, \bibinfo
		{author} {\bibfnamefont {C.}~\bibnamefont {Yang}},\ and\ \bibinfo {author}
		{\bibfnamefont {B.}~\bibnamefont {Judkewitz}},\ }\href
	{https://doi.org/10.1364/OPTICA.1.000227} {\bibfield  {journal} {\bibinfo
			{journal} {Optica}\ }\textbf {\bibinfo {volume} {1}},\ \bibinfo {pages} {227}
		(\bibinfo {year} {2014})}\BibitemShut {NoStop}%
	\bibitem [{\citenamefont {Ma}\ \emph {et~al.}(2014)\citenamefont {Ma},
		\citenamefont {Xu}, \citenamefont {Liu},\ and\ \citenamefont
		{Wang}}]{ma_time-reversed_2014}%
	\BibitemOpen
	\bibfield  {author} {\bibinfo {author} {\bibfnamefont {C.}~\bibnamefont
			{Ma}}, \bibinfo {author} {\bibfnamefont {X.}~\bibnamefont {Xu}}, \bibinfo
		{author} {\bibfnamefont {Y.}~\bibnamefont {Liu}},\ and\ \bibinfo {author}
		{\bibfnamefont {L.~V.}\ \bibnamefont {Wang}},\ }\href
	{https://doi.org/10.1038/nphoton.2014.251} {\bibfield  {journal} {\bibinfo
			{journal} {Nat. Photonics}\ }\textbf {\bibinfo {volume} {8}},\ \bibinfo
		{pages} {931} (\bibinfo {year} {2014})}\BibitemShut {NoStop}%
	\bibitem [{\citenamefont {Ruan}\ \emph {et~al.}(2017)\citenamefont {Ruan},
		\citenamefont {Haber}, \citenamefont {Liu}, \citenamefont {Brake},
		\citenamefont {Kim}, \citenamefont {Berlin},\ and\ \citenamefont
		{Yang}}]{ruan_focusing_2017}%
	\BibitemOpen
	\bibfield  {author} {\bibinfo {author} {\bibfnamefont {H.}~\bibnamefont
			{Ruan}}, \bibinfo {author} {\bibfnamefont {T.}~\bibnamefont {Haber}},
		\bibinfo {author} {\bibfnamefont {Y.}~\bibnamefont {Liu}}, \bibinfo {author}
		{\bibfnamefont {J.}~\bibnamefont {Brake}}, \bibinfo {author} {\bibfnamefont
			{J.}~\bibnamefont {Kim}}, \bibinfo {author} {\bibfnamefont {J.~M.}\
			\bibnamefont {Berlin}},\ and\ \bibinfo {author} {\bibfnamefont
			{C.}~\bibnamefont {Yang}},\ }\href {https://doi.org/10.1364/OPTICA.4.001337}
	{\bibfield  {journal} {\bibinfo  {journal} {Optica}\ }\textbf {\bibinfo
			{volume} {4}},\ \bibinfo {pages} {1337} (\bibinfo {year} {2017})}\BibitemShut
	{NoStop}%
	\bibitem [{\citenamefont {Bouchet}\ \emph
		{et~al.}(2021{\natexlab{b}})\citenamefont {Bouchet}, \citenamefont
		{Seifert},\ and\ \citenamefont {Mosk}}]{bouchet_optimizing_2021}%
	\BibitemOpen
	\bibfield  {author} {\bibinfo {author} {\bibfnamefont {D.}~\bibnamefont
			{Bouchet}}, \bibinfo {author} {\bibfnamefont {J.}~\bibnamefont {Seifert}},\
		and\ \bibinfo {author} {\bibfnamefont {A.~P.}\ \bibnamefont {Mosk}},\ }\href
	{https://www.osapublishing.org/ol/abstract.cfm?uri=ol-46-2-254} {\bibfield
		{journal} {\bibinfo  {journal} {Opt. Lett.}\ }\textbf {\bibinfo {volume}
			{46}},\ \bibinfo {pages} {254} (\bibinfo {year}
		{2021}{\natexlab{b}})}\BibitemShut {NoStop}%
	\bibitem [{\citenamefont {Pappu}\ \emph {et~al.}(2002)\citenamefont {Pappu},
		\citenamefont {Recht}, \citenamefont {Taylor},\ and\ \citenamefont
		{Gershenfeld}}]{pappu_physical_2002}%
	\BibitemOpen
	\bibfield  {author} {\bibinfo {author} {\bibfnamefont {R.}~\bibnamefont
			{Pappu}}, \bibinfo {author} {\bibfnamefont {B.}~\bibnamefont {Recht}},
		\bibinfo {author} {\bibfnamefont {J.}~\bibnamefont {Taylor}},\ and\ \bibinfo
		{author} {\bibfnamefont {N.}~\bibnamefont {Gershenfeld}},\ }\href
	{https://doi.org/10.1126/science.1074376} {\bibfield  {journal} {\bibinfo
			{journal} {Science}\ }\textbf {\bibinfo {volume} {297}},\ \bibinfo {pages}
		{2026} (\bibinfo {year} {2002})}\BibitemShut {NoStop}%
	\bibitem [{\citenamefont {Uppu}\ \emph {et~al.}(2019)\citenamefont {Uppu},
		\citenamefont {Wolterink}, \citenamefont {Goorden}, \citenamefont {Chen},
		\citenamefont {Škorić}, \citenamefont {Mosk},\ and\ \citenamefont
		{Pinkse}}]{uppu_asymmetric_2019}%
	\BibitemOpen
	\bibfield  {author} {\bibinfo {author} {\bibfnamefont {R.}~\bibnamefont
			{Uppu}}, \bibinfo {author} {\bibfnamefont {T.~A.~W.}\ \bibnamefont
			{Wolterink}}, \bibinfo {author} {\bibfnamefont {S.~A.}\ \bibnamefont
			{Goorden}}, \bibinfo {author} {\bibfnamefont {B.}~\bibnamefont {Chen}},
		\bibinfo {author} {\bibfnamefont {B.}~\bibnamefont {Škorić}}, \bibinfo
		{author} {\bibfnamefont {A.~P.}\ \bibnamefont {Mosk}},\ and\ \bibinfo
		{author} {\bibfnamefont {P.~W.~H.}\ \bibnamefont {Pinkse}},\ }\href
	{https://doi.org/10.1088/2058-9565/ab479f} {\bibfield  {journal} {\bibinfo
			{journal} {Quantum Sci. Technol.}\ }\textbf {\bibinfo {volume} {4}},\
		\bibinfo {pages} {045011} (\bibinfo {year} {2019})}\BibitemShut {NoStop}%
	\bibitem [{\citenamefont {Andersen}\ \emph {et~al.}(2016)\citenamefont
		{Andersen}, \citenamefont {Gehring}, \citenamefont {Marquardt},\ and\
		\citenamefont {Leuchs}}]{andersen_30_2016}%
	\BibitemOpen
	\bibfield  {author} {\bibinfo {author} {\bibfnamefont {U.~L.}\ \bibnamefont
			{Andersen}}, \bibinfo {author} {\bibfnamefont {T.}~\bibnamefont {Gehring}},
		\bibinfo {author} {\bibfnamefont {C.}~\bibnamefont {Marquardt}},\ and\
		\bibinfo {author} {\bibfnamefont {G.}~\bibnamefont {Leuchs}},\ }\href
	{https://doi.org/10.1088/0031-8949/91/5/053001} {\bibfield  {journal}
		{\bibinfo  {journal} {Phys. Scr.}\ }\textbf {\bibinfo {volume} {91}},\
		\bibinfo {pages} {053001} (\bibinfo {year} {2016})}\BibitemShut {NoStop}%
	\bibitem [{\citenamefont {Chesi}\ \emph {et~al.}(2018)\citenamefont {Chesi},
		\citenamefont {Olivares},\ and\ \citenamefont
		{Paris}}]{chesi_squeezing-enhanced_2018}%
	\BibitemOpen
	\bibfield  {author} {\bibinfo {author} {\bibfnamefont {G.}~\bibnamefont
			{Chesi}}, \bibinfo {author} {\bibfnamefont {S.}~\bibnamefont {Olivares}},\
		and\ \bibinfo {author} {\bibfnamefont {M.~G.~A.}\ \bibnamefont {Paris}},\
	}\href {https://doi.org/10.1103/PhysRevA.97.032315} {\bibfield  {journal}
		{\bibinfo  {journal} {Phys. Rev. A}\ }\textbf {\bibinfo {volume} {97}},\
		\bibinfo {pages} {032315} (\bibinfo {year} {2018})}\BibitemShut {NoStop}%
	\bibitem [{\citenamefont {Loudon}(2000)}]{loudon_quantum_2000}%
	\BibitemOpen
	\bibfield  {author} {\bibinfo {author} {\bibfnamefont {R.}~\bibnamefont
			{Loudon}},\ }\href@noop {} {\emph {\bibinfo {title} {The {Quantum} {Theory}
				of {Light}}}}\ (\bibinfo  {publisher} {Oxford University Press},\ \bibinfo
	{address} {Oxford},\ \bibinfo {year} {2000})\BibitemShut {NoStop}%
	\bibitem [{\citenamefont {Goodman}(2015)}]{goodman_statistical_2015}%
	\BibitemOpen
	\bibfield  {author} {\bibinfo {author} {\bibfnamefont {J.~W.}\ \bibnamefont
			{Goodman}},\ }\href@noop {} {\emph {\bibinfo {title} {Statistical
				{Optics}}}}\ (\bibinfo  {publisher} {John Wiley \& Sons},\ \bibinfo {address}
	{Hoboken},\ \bibinfo {year} {2015})\BibitemShut {NoStop}%
	\bibitem [{\citenamefont {Pai}\ \emph {et~al.}(2020)\citenamefont {Pai},
		\citenamefont {Bosch},\ and\ \citenamefont {Mosk}}]{pai_optical_2020}%
	\BibitemOpen
	\bibfield  {author} {\bibinfo {author} {\bibfnamefont {P.}~\bibnamefont
			{Pai}}, \bibinfo {author} {\bibfnamefont {J.}~\bibnamefont {Bosch}},\ and\
		\bibinfo {author} {\bibfnamefont {A.~P.}\ \bibnamefont {Mosk}},\ }\href
	{https://doi.org/10.1364/OSAC.384832} {\bibfield  {journal} {\bibinfo
			{journal} {OSA Continuum}\ }\textbf {\bibinfo {volume} {3}},\ \bibinfo
		{pages} {637} (\bibinfo {year} {2020})}\BibitemShut {NoStop}%
	\bibitem [{\citenamefont {Mirhosseini}\ \emph {et~al.}(2013)\citenamefont
		{Mirhosseini}, \citenamefont {Magaña-Loaiza}, \citenamefont {Chen},
		\citenamefont {Rodenburg}, \citenamefont {Malik},\ and\ \citenamefont
		{Boyd}}]{mirhosseini_rapid_2013}%
	\BibitemOpen
	\bibfield  {author} {\bibinfo {author} {\bibfnamefont {M.}~\bibnamefont
			{Mirhosseini}}, \bibinfo {author} {\bibfnamefont {O.~S.}\ \bibnamefont
			{Magaña-Loaiza}}, \bibinfo {author} {\bibfnamefont {C.}~\bibnamefont
			{Chen}}, \bibinfo {author} {\bibfnamefont {B.}~\bibnamefont {Rodenburg}},
		\bibinfo {author} {\bibfnamefont {M.}~\bibnamefont {Malik}},\ and\ \bibinfo
		{author} {\bibfnamefont {R.~W.}\ \bibnamefont {Boyd}},\ }\href
	{https://doi.org/10.1364/OE.21.030196} {\bibfield  {journal} {\bibinfo
			{journal} {Opt. Express}\ }\textbf {\bibinfo {volume} {21}},\ \bibinfo
		{pages} {30196} (\bibinfo {year} {2013})}\BibitemShut {NoStop}%
\end{thebibliography}

\begin{thebibliography}{10}%
	\makeatletter
	\providecommand \@ifxundefined [1]{%
		\@ifx{#1\undefined}
	}%
	\providecommand \@ifnum [1]{%
		\ifnum #1\expandafter \@firstoftwo
		\else \expandafter \@secondoftwo
		\fi
	}%
	\providecommand \@ifx [1]{%
		\ifx #1\expandafter \@firstoftwo
		\else \expandafter \@secondoftwo
		\fi
	}%
	\providecommand \natexlab [1]{#1}%
	\providecommand \enquote  [1]{``#1''}%
	\providecommand \bibnamefont  [1]{#1}%
	\providecommand \bibfnamefont [1]{#1}%
	\providecommand \citenamefont [1]{#1}%
	\providecommand \href@noop [0]{\@secondoftwo}%
	\providecommand \href [0]{\begingroup \@sanitize@url \@href}%
	\providecommand \@href[1]{\@@startlink{#1}\@@href}%
	\providecommand \@@href[1]{\endgroup#1\@@endlink}%
	\providecommand \@sanitize@url [0]{\catcode `\\12\catcode `\$12\catcode
		`\&12\catcode `\#12\catcode `\^12\catcode `\_12\catcode `\%12\relax}%
	\providecommand \@@startlink[1]{}%
	\providecommand \@@endlink[0]{}%
	\providecommand \url  [0]{\begingroup\@sanitize@url \@url }%
	\providecommand \@url [1]{\endgroup\@href {#1}{\urlprefix }}%
	\providecommand \urlprefix  [0]{URL }%
	\providecommand \Eprint [0]{\href }%
	\providecommand \doibase [0]{https://doi.org/}%
	\providecommand \selectlanguage [0]{\@gobble}%
	\providecommand \bibinfo  [0]{\@secondoftwo}%
	\providecommand \bibfield  [0]{\@secondoftwo}%
	\providecommand \translation [1]{[#1]}%
	\providecommand \BibitemOpen [0]{}%
	\providecommand \bibitemStop [0]{}%
	\providecommand \bibitemNoStop [0]{.\EOS\space}%
	\providecommand \EOS [0]{\spacefactor3000\relax}%
	\providecommand \BibitemShut  [1]{\csname bibitem#1\endcsname}%
	\let\auto@bib@innerbib\@empty
	%</preamble>
	\bibitem [{\citenamefont {Loudon}(2000)}]{loudon_quantum_2000_2}%
	\BibitemOpen
	\bibfield  {author} {\bibinfo {author} {\bibfnamefont {R.}~\bibnamefont
			{Loudon}},\ }\href@noop {} {\emph {\bibinfo {title} {The {Quantum} {Theory}
				of {Light}}}}\ (\bibinfo  {publisher} {Oxford University Press},\ \bibinfo
	{address} {Oxford},\ \bibinfo {year} {2000})\BibitemShut {NoStop}%
	\bibitem [{\citenamefont {Pai}\ \emph {et~al.}(2021)\citenamefont {Pai},
		\citenamefont {Bosch}, \citenamefont {Kühmayer}, \citenamefont {Rotter},\
		and\ \citenamefont {Mosk}}]{pai_scattering_2021_2}%
	\BibitemOpen
	\bibfield  {author} {\bibinfo {author} {\bibfnamefont {P.}~\bibnamefont
			{Pai}}, \bibinfo {author} {\bibfnamefont {J.}~\bibnamefont {Bosch}}, \bibinfo
		{author} {\bibfnamefont {M.}~\bibnamefont {Kühmayer}}, \bibinfo {author}
		{\bibfnamefont {S.}~\bibnamefont {Rotter}},\ and\ \bibinfo {author}
		{\bibfnamefont {A.~P.}\ \bibnamefont {Mosk}},\ }\href
	{https://doi.org/10.1038/s41566-021-00789-9} {\bibfield  {journal} {\bibinfo
			{journal} {Nat. Photonics}\ }\textbf {\bibinfo {volume} {15}},\ \bibinfo
		{pages} {431} (\bibinfo {year} {2021})}\BibitemShut {NoStop}%
	\bibitem [{\citenamefont {Goodman}(2015)}]{goodman_statistical_2015_2}%
	\BibitemOpen
	\bibfield  {author} {\bibinfo {author} {\bibfnamefont {J.~W.}\ \bibnamefont
			{Goodman}},\ }\href@noop {} {\emph {\bibinfo {title} {Statistical
				{Optics}}}}\ (\bibinfo  {publisher} {John Wiley \& Sons},\ \bibinfo {address}
	{Hoboken},\ \bibinfo {year} {2015})\BibitemShut {NoStop}%
	\bibitem [{\citenamefont {Trees}\ \emph {et~al.}(2013)\citenamefont {Trees},
		\citenamefont {Bell},\ and\ \citenamefont {Tian}}]{trees_detection_2013_2}%
	\BibitemOpen
	\bibfield  {author} {\bibinfo {author} {\bibfnamefont {H.~L.~V.}\
			\bibnamefont {Trees}}, \bibinfo {author} {\bibfnamefont {K.~L.}\ \bibnamefont
			{Bell}},\ and\ \bibinfo {author} {\bibfnamefont {Z.}~\bibnamefont {Tian}},\
	}\href@noop {} {\emph {\bibinfo {title} {Detection {Estimation} and
				{Modulation} {Theory}, {Part} {I}}}}\ (\bibinfo  {publisher} {John Wiley \&
		Sons},\ \bibinfo {address} {Hoboken},\ \bibinfo {year} {2013})\BibitemShut
	{NoStop}%
	\bibitem [{\citenamefont {Lee}(1974)}]{lee_binary_1974_2}%
	\BibitemOpen
	\bibfield  {author} {\bibinfo {author} {\bibfnamefont {W.-H.}\ \bibnamefont
			{Lee}},\ }\href {https://doi.org/10.1364/AO.13.001677} {\bibfield  {journal}
		{\bibinfo  {journal} {Appl. Opt.}\ }\textbf {\bibinfo {volume} {13}},\
		\bibinfo {pages} {1677} (\bibinfo {year} {1974})}\BibitemShut {NoStop}%
	\bibitem [{\citenamefont {Cuche}\ \emph {et~al.}(2000)\citenamefont {Cuche},
		\citenamefont {Marquet},\ and\ \citenamefont
		{Depeursinge}}]{cuche_spatial_2000_2}%
	\BibitemOpen
	\bibfield  {author} {\bibinfo {author} {\bibfnamefont {E.}~\bibnamefont
			{Cuche}}, \bibinfo {author} {\bibfnamefont {P.}~\bibnamefont {Marquet}},\
		and\ \bibinfo {author} {\bibfnamefont {C.}~\bibnamefont {Depeursinge}},\
	}\href {https://doi.org/10.1364/AO.39.004070} {\bibfield  {journal} {\bibinfo
			{journal} {Appl. Opt.}\ }\textbf {\bibinfo {volume} {39}},\ \bibinfo {pages}
		{4070} (\bibinfo {year} {2000})}\BibitemShut {NoStop}%
	\bibitem [{\citenamefont {Popoff}\ \emph {et~al.}(2010)\citenamefont {Popoff},
		\citenamefont {Lerosey}, \citenamefont {Carminati}, \citenamefont {Fink},
		\citenamefont {Boccara},\ and\ \citenamefont
		{Gigan}}]{popoff_measuring_2010_2}%
	\BibitemOpen
	\bibfield  {author} {\bibinfo {author} {\bibfnamefont {S.~M.}\ \bibnamefont
			{Popoff}}, \bibinfo {author} {\bibfnamefont {G.}~\bibnamefont {Lerosey}},
		\bibinfo {author} {\bibfnamefont {R.}~\bibnamefont {Carminati}}, \bibinfo
		{author} {\bibfnamefont {M.}~\bibnamefont {Fink}}, \bibinfo {author}
		{\bibfnamefont {A.~C.}\ \bibnamefont {Boccara}},\ and\ \bibinfo {author}
		{\bibfnamefont {S.}~\bibnamefont {Gigan}},\ }\href
	{https://doi.org/10.1103/PhysRevLett.104.100601} {\bibfield  {journal}
		{\bibinfo  {journal} {Phys. Rev. Lett.}\ }\textbf {\bibinfo {volume} {104}},\
		\bibinfo {pages} {100601} (\bibinfo {year} {2010})}\BibitemShut {NoStop}%
	\bibitem [{\citenamefont {Pai}\ \emph {et~al.}(2020)\citenamefont {Pai},
		\citenamefont {Bosch},\ and\ \citenamefont {Mosk}}]{pai_optical_2020_2}%
	\BibitemOpen
	\bibfield  {author} {\bibinfo {author} {\bibfnamefont {P.}~\bibnamefont
			{Pai}}, \bibinfo {author} {\bibfnamefont {J.}~\bibnamefont {Bosch}},\ and\
		\bibinfo {author} {\bibfnamefont {A.~P.}\ \bibnamefont {Mosk}},\ }\href
	{https://doi.org/10.1364/OSAC.384832} {\bibfield  {journal} {\bibinfo
			{journal} {OSA Continuum}\ }\textbf {\bibinfo {volume} {3}},\ \bibinfo
		{pages} {637} (\bibinfo {year} {2020})}\BibitemShut {NoStop}%
	\bibitem [{\citenamefont {Mirhosseini}\ \emph {et~al.}(2013)\citenamefont
		{Mirhosseini}, \citenamefont {Magaña-Loaiza}, \citenamefont {Chen},
		\citenamefont {Rodenburg}, \citenamefont {Malik},\ and\ \citenamefont
		{Boyd}}]{mirhosseini_rapid_2013_2}%
	\BibitemOpen
	\bibfield  {author} {\bibinfo {author} {\bibfnamefont {M.}~\bibnamefont
			{Mirhosseini}}, \bibinfo {author} {\bibfnamefont {O.~S.}\ \bibnamefont
			{Magaña-Loaiza}}, \bibinfo {author} {\bibfnamefont {C.}~\bibnamefont
			{Chen}}, \bibinfo {author} {\bibfnamefont {B.}~\bibnamefont {Rodenburg}},
		\bibinfo {author} {\bibfnamefont {M.}~\bibnamefont {Malik}},\ and\ \bibinfo
		{author} {\bibfnamefont {R.~W.}\ \bibnamefont {Boyd}},\ }\href
	{https://doi.org/10.1364/OE.21.030196} {\bibfield  {journal} {\bibinfo
			{journal} {Opt. Express}\ }\textbf {\bibinfo {volume} {21}},\ \bibinfo
		{pages} {30196} (\bibinfo {year} {2013})}\BibitemShut {NoStop}%
	\bibitem [{\citenamefont {Mosk}\ \emph {et~al.}(2012)\citenamefont {Mosk},
		\citenamefont {Lagendijk}, \citenamefont {Lerosey},\ and\ \citenamefont
		{Fink}}]{mosk_controlling_2012_2}%
	\BibitemOpen
	\bibfield  {author} {\bibinfo {author} {\bibfnamefont {A.~P.}\ \bibnamefont
			{Mosk}}, \bibinfo {author} {\bibfnamefont {A.}~\bibnamefont {Lagendijk}},
		\bibinfo {author} {\bibfnamefont {G.}~\bibnamefont {Lerosey}},\ and\ \bibinfo
		{author} {\bibfnamefont {M.}~\bibnamefont {Fink}},\ }\href
	{https://doi.org/10.1038/nphoton.2012.88} {\bibfield  {journal} {\bibinfo
			{journal} {Nat. Photonics}\ }\textbf {\bibinfo {volume} {6}},\ \bibinfo
		{pages} {283} (\bibinfo {year} {2012})}\BibitemShut {NoStop}%
\end{thebibliography}
\end{document}